\documentclass[manuscript]{acmart} % tosem
\usepackage{arydshln}
\usepackage[font=small,labelfont=bf,tableposition=top]{caption}
\usepackage{hyperref}
\usepackage{boldline}
\usepackage{color,colortbl}
\usepackage{bigstrut}
\usepackage[ruled]{algorithm2e}
\usepackage{amsmath,amsfonts}
\usepackage{amsmath}
\usepackage{array}
\usepackage{algorithmic}
\usepackage{balance}
\usepackage{blindtext}
\usepackage{booktabs}
\usepackage{caption}
\usepackage{hyperref}
\usepackage[noabbrev]{cleveref}
\usepackage{comment}
\usepackage{enumitem}
\usepackage{eqparbox}
\usepackage{fancybox}
\usepackage{fancyvrb}
\usepackage{framed}
\usepackage{graphicx}
\usepackage{ifthen}
\usepackage{listings}
\usepackage{mathrsfs}
\usepackage{mdwmath}
\usepackage{mdwtab}
\usepackage{multirow}
\usepackage{pifont}
\usepackage{stfloats}
\usepackage{textcomp}
\usepackage{url}
\usepackage{xcolor}
\usepackage{xspace}
\usepackage[normalem]{ulem}
\usepackage[framemethod=TikZ]{mdframed}
\usepackage{caption}
\usepackage{subcaption}
\usepackage{tabularx}
\usepackage{tcolorbox}
\tcbuselibrary{listings,skins}
\usepackage{mathtools}
\usepackage{blindtext}
\usepackage{lstautogobble}
\usepackage{pifont}

\usepackage{xspace}
\xspace%

\usepackage{tcolorbox}

\newtcbox{\inlinebox}[1][]{enhanced,
 box align=base,
 nobeforeafter,
 colback=blueish,
 size=small,
 left=0pt,
 right=0pt,
 boxsep=2pt,
 #1}

\newcommand{\lessons}[1]{
    \begin{lesson}
    \small
        Observation~#1
    \end{lesson}
}

\usepackage{cleveref}

\renewcommand{\cref}[1]{\Cref{#1}}

\newcommand{\rom}[1]{\uppercase\expandafter{\romannumeral #1\relax}}

\definecolor{gray50}{gray}{.5}
\definecolor{gray40}{gray}{.6}
\definecolor{gray30}{gray}{.7}
\definecolor{gray20}{gray}{.8}
\definecolor{gray10}{gray}{.9}
\definecolor{gray05}{gray}{.95}
\definecolor{gray01}{gray}{.97}

\newlength\Linewidth
\def\findlength{\setlength\Linewidth\linewidth
\addtolength\Linewidth{-4\fboxrule}
\addtolength\Linewidth{-3\fboxsep}
}
\newenvironment{examplebox}{\par\begingroup
  \setlength{\fboxsep}{3pt}\findlength
  \setbox0=\vbox\bgroup\noindent
  \hsize=0.90\linewidth
  \begin{minipage}{0.90\linewidth}\normalsize}
    {\end{minipage}\egroup
    \textcolor{gray20}{\fboxsep1.2pt\fbox
     {\fboxsep5pt\colorbox{gray05}{\normalcolor\box0}}}
    \endgroup\par\noindent
    \normalcolor\ignorespacesafterend}

\usepackage{tikz}
\usetikzlibrary{arrows.meta, shapes, positioning, calc}
\usepackage{mdframed}

\usetikzlibrary{shadows}
\usepackage{graphics}
\newmdenv[
    tikzsetting= {fill=blueish},
    skipabove=0.33em,
    skipbelow=0.33em,
    linewidth=1pt,
    innerleftmargin=4pt,
    innerrightmargin=4pt,
    innertopmargin=2pt,
    innerbottommargin=2pt,
    linecolor=gray95,
    roundcorner=2pt, 
    shadow=true,
    shadowsize=4pt,
    shadowcolor=gray95
]{questionbox}

\newmdenv[
    tikzsetting= {fill=greenish},
    skipabove=0.33em,
    skipbelow=0.33em,
    linewidth=1pt,
    innerleftmargin=4pt,
    innerrightmargin=4pt,
    innertopmargin=2pt,
    innerbottommargin=2pt,
    linecolor=gray95,
    roundcorner=2pt, 
    shadow=true,
    shadowsize=4pt,
    shadowcolor=gray95
]{answerbox}

\newmdenv[
    skipabove=0.33em,
    skipbelow=0.33em,
    innerleftmargin=4pt,
    innerrightmargin=4pt,
    innertopmargin=2pt,
    innerbottommargin=2pt,
]{lessonbox}

\usepackage{tikz}

\usepackage{tabularx}

\newenvironment{lesson}
{
    \begin{lessonbox}
}
{\end{lessonbox}}

\newenvironment{result}
{\begin{answerbox}}
{\end{answerbox}}

\newenvironment{question}
{\begin{questionbox}}
{\end{questionbox}}

\definecolor{javared}{rgb}{0.6,0,0} % for strings
\definecolor{javagreen}{rgb}{0.25,0.5,0.35} % comments
\definecolor{javapurple}{rgb}{0.5,0,0.35} % keywords
\definecolor{javadocblue}{rgb}{0.25,0.35,0.75} % javadoc

\lstdefinestyle{basejava}{
  language=java,
  showstringspaces=false,
  basicstyle=\scriptsize\ttfamily,
  keywordstyle=\bfseries\color{javapurple},
  commentstyle=\itshape\blue,
  identifierstyle=\blue,
  frame=none,
  backgroundcolor=\color{white},
}

\lstdefinestyle{CustomJava}{
  belowcaptionskip=\baselineskip,
  breaklines=true,
  xleftmargin=\parindent,
  language=java,
  showstringspaces=false,
  basicstyle=\scriptsize\ttfamily,
  keywordstyle=\bfseries\color{javapurple},
  commentstyle=\itshape\blue,
  identifierstyle=\blue,
  belowskip=1pt,
  frame=shadowbox,
  backgroundcolor=\color{gray01},
  gobble=0
}

\usepackage{tcolorbox}
\tcbuselibrary{listingsutf8}  % This library adds utf8 support for listings

\newtcblisting{mylisting}[2][]{
    arc=3mm,
    listing only, 
    listing style=codit,
    title=#2,
    #1
    }

\newcommand\blue[1]{\textcolor[rgb]{0.00,0.00,1.00}{{#1}}}

\definecolor{blueish}{RGB}{252, 252, 255}
\definecolor{greenish}{RGB}{252, 255, 252}
\definecolor{redish}{RGB}{255, 250, 250}

\definecolor{gray05}{gray}{0.95}
\definecolor{gray08}{gray}{0.92}
\definecolor{gray10}{gray}{0.90}
\definecolor{gray12}{gray}{0.88}
\definecolor{gray15}{gray}{0.85}
\definecolor{gray18}{gray}{0.82}
\definecolor{gray20}{gray}{0.80}
\definecolor{gray25}{gray}{0.75}
\definecolor{gray30}{gray}{0.70}
\definecolor{gray35}{gray}{0.65}
\definecolor{gray40}{gray}{0.60}
\definecolor{gray45}{gray}{0.55}
\definecolor{gray50}{gray}{0.50}
\definecolor{gray55}{gray}{0.45}
\definecolor{gray60}{gray}{0.40}
\definecolor{gray65}{gray}{0.35}
\definecolor{gray70}{gray}{0.30}
\definecolor{gray75}{gray}{0.25}
\definecolor{gray80}{gray}{0.20}
\definecolor{gray85}{gray}{0.15}
\definecolor{gray90}{gray}{0.10}
\definecolor{gray95}{gray}{0.05}

\definecolor{fstarid}{rgb}{0.28,0.07,0.07}

\definecolor{addition}{rgb}{0,0.1,0.5}
\definecolor{dkblue}{rgb}{0,0.0,0.7}
\definecolor{dkgreen}{rgb}{0,0.4,0}
\definecolor{dkred}{rgb}{0.6,0,0}
\definecolor{dkpurple}{rgb}{0.55,0,0.75}
\definecolor{purple}{rgb}{0.69,0,.87}
\definecolor{olive}{rgb}{0.4, 0.4, 0.0}
\definecolor{teal}{rgb}{0.0,0.4,0.4}
\definecolor{azure}{rgb}{0.0, 0.5, 1.0}
\definecolor{gray}{rgb}{0.5, 0.5, 0.5}
\definecolor{dkgrey}{rgb}{0.2, 0.2, 0.2}
\definecolor{lilac}{rgb}{0.70, 0.04, 0.08}
\definecolor{applegreen}{rgb}{0.55, 0.71, 0.0}

\definecolor{step1}{HTML}{CC0066}
\definecolor{step2}{HTML}{CC6600}
\definecolor{step3}{HTML}{440077}
\definecolor{step4}{HTML}{007700}
\definecolor{step5}{HTML}{0000FF}

\usepackage[turnon]{notes}

\usepackage{amsmath}
\usepackage{amsthm}
\usepackage{mathrsfs}
\usepackage{makecell}

\usepackage{balance}
\newcounter{o}
\AtBeginDocument{%
  }

\setcopyright{acmlicensed}
\copyrightyear{2018}
\acmYear{2018}
\acmDOI{XXXXXXX.XXXXXXX}

\acmConference[Conference acronym 'XX]{Make sure to enter the correct
  conference title from your rights confirmation email}{June 03--05,
  2018}{Woodstock, NY}
\acmISBN{978-1-4503-XXXX-X/18/06}

\begin{document}

%\title{MetaCodeAgent: Coding with LLMs using Metamorphic Specification Mutation}
\title{LLM Assisted Coding with Metamorphic Specification Mutation Agent}
\author{Mostafijur Rahman Akhond}
\email{mostafij@yorku.ca}
\orcid{0000-0003-4902-4858}
\affiliation{%
  \institution{York University}
%   % \city{}
  \country{Canada}
}

% \author{Ran Bi}
% \email{ryan1212@yorku.ca}
% \orcid{}
% \affiliation{%
%   \institution{York University}  
%   \country{Canada}}

% \author{Saikat Chakraborty }
% \affiliation{%
%   \institution{Microsoft Research}
%   % \city{Hekla}
% \email{saikatc@microsoft.com}

\author{Gias Uddin}
\email{guddin@yorku.ca}
\orcid{0000-0003-1376-095X}
\affiliation{%
  \institution{York University}  
  \country{Canada}}

\renewcommand{\shortauthors}{Akhond and Uddin}

\begin{abstract}

   Metamorphic Relations (MRs) serve as a foundational mechanism for generating semantically equivalent mutations. Software engineering has advanced significantly in recent years with the advent of Large Language Models (LLMs). However, the reliability of LLMs in software engineering is often compromised by ambiguities and inconsistencies due to improper user specification. To address this challenge, we present \textit{CodeMetaAgent (CMA)}, a metamorphic relation–driven LLM-agent that systematically refines task specifications and generates semantically constrained test cases. Our proposed framework uses MRs with LLMs to improve generation consistency and reduce variability caused by specifications, unlike the traditional use of MRs as post validations. Our framework has been experimented on the HumanEval-Pro, MBPP-Pro, and SWE-Bench\_Lite datasets with the GPT-4o, Mistral Large, GPT-OSS, and Qwen3-Coder models. It improved code generation accuracy by up to 17\% and gained code coverage by up to 99.81\%. These results show that metamorphic relations can be a simple but effective guide in assisting LLM-based software development.
\end{abstract}




\received{10 July 2025}
\received[revised]{\textsc{20 July 2025}}
\received[accepted]{5 Dec 2025}

\maketitle

\section{Introduction}

% Large Language Models (LLMs) have revolutionized software development by providing powerful tools for tasks such as code generation \cite{jin2024can, barke2023grounded, chen2021evaluating}, test case generation \cite{chen2024chatunitest, alshahwan2024observation, schafer2023empirical}, and bug fixing \cite{jimenez2023swe, xia2023automated}. While LLMs have demonstrated impressive capabilities in understanding and generating code, the issue of hallucination remains a significant challenge \cite{liu2024exploring, wu2025detecting}. Hallucination refers to the generation of incorrect or misleading code by LLMs, which can lead to errors, inconsistencies, and inefficiencies in software development processes \cite{huang2025survey}. Researchers have explored various techniques to mitigate hallucinations in LLMs, such as prompt engineering, fine-tuning, and the integration of external knowledge bases. However, these approaches often fall short in addressing the inherent complexity and non-determinism associated with coding tasks.

Metamorphic Relations (MRs) provide a structured framework for the transformation of inputs or specifications\cite{segura2016survey, liu2012new}. MRs capture semantic invariants that should hold across multiple representations of a task. It enables the validation or refinement of software even in the absence of ground-truth oracles. This property makes MRs especially useful for Large Language Models (LLMs), where code generation \cite{jin2024can, barke2023grounded, chen2021evaluating}, bug fixing \cite{jimenez2023swe, xia2023automated}, and test synthesis \cite{chen2024chatunitest, alshahwan2024observation, schafer2023empirical} are very sensitive to changes from the user descriptions. LLMs can produce more accurate, consistent, and robust outputs across a wide range of software engineering tasks by using problem specifications to logically equivalent mutations.

Large Language Models (LLMs) like GPT ~\cite{achiam2023gpt}, Qwen ~\cite{yang2025qwen3}, and Mistral ~\cite{mistral_large} have demonstrated that they can translate natural language specifications into executable code, produce test cases corresponding to the source codes, and identify and resolve bugs from the program ~\cite{chen2021evaluating, yu2024humaneval, jin2024can}.  Even though these models work well, the outputs they give can be very different depending on how the input is presented or what assumptions are made.  Minor linguistic variations can result in divergent reasoning pathways and lead to inconsistencies \cite{Chen2024NLPerturbatorST, Shen2023InCW}. Recent prompt-based refinement or retrieval-augmented methods try to fix this problem, but they struggle to make the model robust against ambiguous specifications. This gap shows that we need a systematic way for LLMs to think about multiple aligned ways to solve a problem, instead of just one static prompt.

We present \textit{CodeMetaAgent (CMA)}, a metamorphic relation–driven agent that enhances LLM-based software development through systematic specification refinement and validation. Instead of using a single static prompt, CMA uses metamorphic relations as a semantic operator that creates and evaluates mutations. This enables LLMs to explore different ways of reasoning, which makes different Software Engineering (SE) related tasks, for example, code generation, or test case synthesis, more robust. By integrating transformation, validation, generation, and execution within a unified pipeline.

In the \textit{CMA} framework, metamorphic relations are facilitated to enhance both code generation and test case synthesis. For the code generation tasks, MRs are applied to reformulate the original problem description into multiple semantically equivalent variants. These variants expose several reasoning perspectives, enabling LLMs to generalize beyond a single prompt and produce more accurate and reliable implementations. For the test cases generation, the relations encode expected behavioral properties over input–output transformations. This allows the automatic derivation of test cases covering divergent conditions and edge cases.  CMA sets up a way to improve robustness and coverage in LLM-assisted software development by making it easier to generate and validate these two types of mutations.

In this work, we make three key contributions. First, we introduce \textit{CMA}, a unified agentic framework that leverages metamorphic relations as proactive semantic operators, transforming MRs from post-hoc validation mechanisms to core drivers of code and test generation. Second, we formalize the role of metamorphic relations across two fundamental dimensions of software engineering—code generation and test case generation, defining task-specific transformations that enable systematic refinement with semantic alignment. Third, we show that our framework significantly improves code generation accuracy (up to 17\% improvement), and achieves test coverage as high as 99.81\% through extensive experiments on three widely used benchmarks (HumanEval Pro, MBPP Pro, and SWE-Bench-Lite) using state-of-the-art LLMs. These results establish metamorphic relation–guided agentic reasoning as a principled approach for improving the robustness, reliability, and consistency of LLM-based software development.

Ensuring semantic robustness in LLM-assisted software engineering is increasingly critical as these models are integrated into practical development workflows. This research investigates the systematic enhancement of reasoning reliability, accuracy, and comprehensiveness of LLMs through metamorphic relations in essential software engineering tasks. We focus on two top-level research questions: (1) can metamorphic relations make LLM better at generating code and fixing bugs, and (2) can they make LLM-generated test cases more useful and cover more edge cases? The rest of the paper discusses the Background, CMA framework, the experimental setup, and the empirical evaluation to demonstrate its effectiveness.

\section{Background \& Related Work}

\subsection{Metamorphic Relations in Software Engineering}

A \textit{metamorphic relation} (MR) is defined as how the output of a function ought to change when its input is systematically altered \cite{zhou2018metamorphic, mayer2006empirical}. Instead of verifying whether a single output is correct in isolation, an MR defines a relationship between the outputs corresponding to inputs. Consider the function \( f(x) = x^2 \) as an example. If we double the input from \( x \) to \( 2x \), the output should increase by a factor of four, since \( f(2x) = (2x)^2 = 4x^2 = 4f(x) \). This consistent behavior under a specific input transformation is a metamorphic relation.

Metamorphic relations (MRs) have played a central role in improving software engineering activities. The application of MRs in software engineering ranges from reliability, testing, and verification \cite{chen2020metamorphic, liu2012new, zhou2018metamorphic}. Traditionally, MRs were employed as part of metamorphic testing (MT), a methodology that validates program correctness without requiring test oracles. By expressing expected relationships between multiple executions of a program with transformed inputs, MRs enable testers to detect inconsistencies that indicate latent errors. Beyond their role in testing, recent work has explored MRs as semantic specifications \cite{xu2024mr, wu2025detecting}. This enables automated reasoning about the software behavior in scenarios where explicit expected outputs are not available. 

The adoption of metamorphic relations in software testing workflows has gained momentum in recent years. Xu et al. \cite{xu2024mr} proposed a framework for automatically discovering MRs to support black-box testing. In another work, Hoard et al. \cite{hoard2025acceptance} investigated developer acceptance of MR-based testing in industrial settings. They emphasized the need for agents in automation. These studies show that MRs improve fault detection and developer productivity. However, they still treat MRs as secondary validation mechanisms, rather than primary operators that actively drive the software development process. In contrast, in this work, we presented MRs as semantic operators embedded within an agentic framework to enhance the reasoning capabilities of Large Language Models.

\subsection{LLMs as Software Engineering Assistants}

Large Language Models (LLMs) have become a useful assistant in automating software engineering activities. It helps developers in code generation, refactoring, bug fixing, and testing. Trained on vast collections of source code and natural language descriptions, LLMs like GPT models, Code LLaMA, Mistral, and Qwen Coder are capable of translating problem statements into syntactically and functionally valid programs \cite{chen2021evaluating, austin2021program}. LLMs also help formulate the validation of programs using different testing suites \cite{ouedraogo2024large, chen2021evaluating, jin2024can}.   

However, despite their impressive performance, LLMs face significant challenges. They may produce code that is syntactically correct but semantically flawed, overfit to familiar patterns from training data, or fail to generalize when problem descriptions are rephrased \cite{xu2022systematic}. Furthermore, conventional evaluation approaches that rely solely on static test suites may miss subtle logical errors, leading to incorrect or inconsistent outputs. These limitations have motivated the integration of metamorphic testing principles into LLM-based code generation pipelines, enhancing both the generation process and the evaluation of the resulting code \cite{wu2025detecting,chen2020metamorphic}.

For example, Wu et al. \cite{wu2025detecting} demonstrated that LLMs can suffer from hallucinations when developing source code from user specifications. They proposed a metamorphic testing-based question answering process to detect and reduce factual hallucinations. By defining predictable relationships between the inputs and outputs, their system generated a more reliable mechanism for code synthesis.

\subsection{LLM Agents in Software Engineering}

An LLM-agent in software engineering is an autonomous framework that leverages large language models (LLMs) to perform complex software engineering tasks. Traditionally, agents are termed an a intellignet entity falititated by perceiving, reasoning and applying actions to achieve specific goals \cite{Wang2023ASO}. When agents applied to LLms, it combines the generative capabilities of LLMs with modular processes that allow the agent to decompose problems, generate intermediate outputs, validate the solutions, and refine if necessary \cite{Liu2024LargeLM, Zhang2024CodeAgentEC}.

% LLM Agents involved in software engineering can be broadly categorized into five types: 1) Simple Reflex agents 2) Model-based agents 3) Goal-based agents 4) Utility-based agents and 5) Learning agents \cite{bytebytego-ai-agents2025}. 

% The Simple Reflex agents are considered as the most basic type of agents. 

The rise of LLM-agents introduces a major shift in automated software engineering. Unlike single-pass prompting, agentic frameworks support multi-step reasoning \cite{yang2024swe, qian2023chatdev}. They enable task decomposition, intermediate solution generation, reflection, and strategy adaptation. This enables LLMs to surpass the limitations of static code generation. They support multi-step debugging, automated test synthesis, and specification refinement.

LLM agents typically adopt a modular architecture, whereby dedicated components collaborate to fulfill distinct roles. For instance, frameworks such as ChatDev \cite{qian2023chatdev} and SWE‑Agent \cite{yang2024swe} orchestrate multiple LLM-backed modules to generate software artifacts, evaluate progress, and iterate until the task is complete. Another exemplar is OpenHands, an open-source AI-coding-agent platform that enables LLM-based agents to write code, run commands, browse documentation, and manage entire repositories based on natural-language specifications \cite{wang2024openhands}. The agentic design thus enables LLMs to emulate human-like software development workflows—planning, execution, feedback, and revision—while scaling across tasks and domains.

% different types of LLM agents Goal based - Utility based - Learning agents - Simple Reflex - Model-based

\section{Methodology: CMA}
\label{sec:metacodeagent}

We propose \textit{CMA}, an MR-driven LLM agent for SE designed to enhance Large Language Model (LLM)-based software development tasks, including code generation, bug fixing, and metamorphic test case generation. CMA backed metamorphic relations as proactive semantic operators that systematically transform and refine both task specifications and test inputs. These MR-guided transformations enable LLMs to reason over multiple semantically equivalent representations, thereby increasing robustness, improving correctness, and enhancing test coverage.

\begin{figure}[!t]
    \centering
    \includegraphics[width=\linewidth]{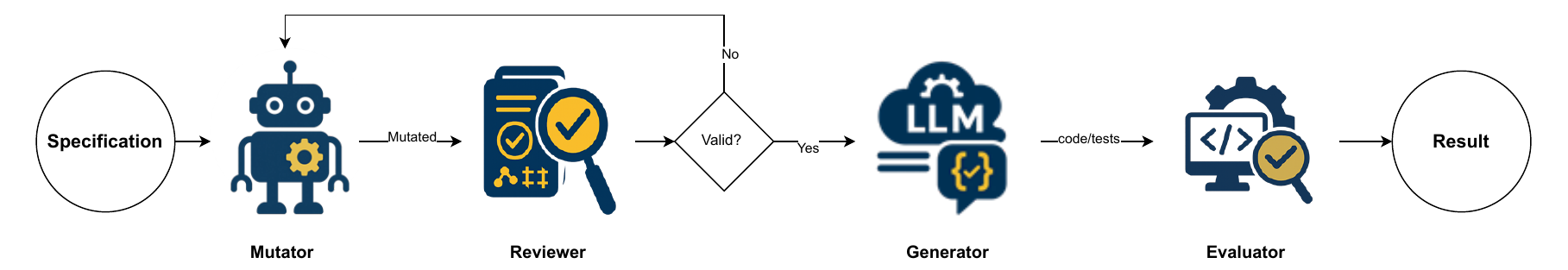}
    \caption{Workflow of the \textsc{CMA} framework.}
    \label{fig:metacodeagent_framework}
\end{figure}

As illustrated in Figure~\ref{fig:metacodeagent_framework}, the framework consists of four autonomous modules—the \textbf{Mutator}, \textbf{Reviewer}, \textbf{Generator}, and \textbf{Evaluator}. The Mutator generates MR-based transformations, the Reviewer validates their semantic fidelity, the Generator prompts LLM to generate code, patches, or test cases, and the Evaluator executes and assesses the generated outputs. Together, these components form an iterative pipeline that enables MRs as active drivers of LLM reasoning.

At a high level, CMA operates across three pipelines: 
(1) \textit{code generation}, where problem descriptions are transformed using MRs and used to prompt the LLM to synthesize solutions; 
(2) \textit{bug fixing}, where issue reports undergo MR-guided transformation to improve clarity and reasoning quality of the generated patches; and 
(3) \textit{test case generation}, where oracle test inputs are metamorphically transformed into variants that assess the robustness and behavioral consistency of the generated code.

\begin{figure*}[t]
    \centering
    \includegraphics[width=\linewidth]{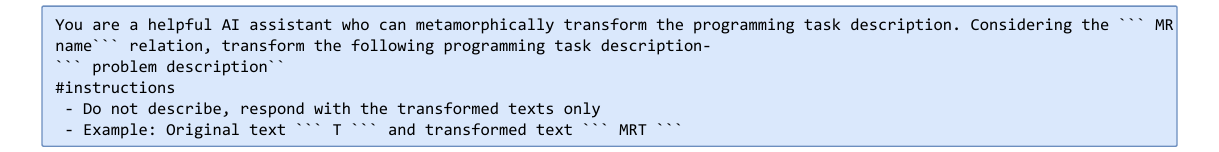}
    \caption{Prompt used to generate metamorphically transformed problem description}
    \label{fig:prompt_mr}
\end{figure*}

\begin{figure*}[t]
    \centering
    \includegraphics[width=\linewidth]{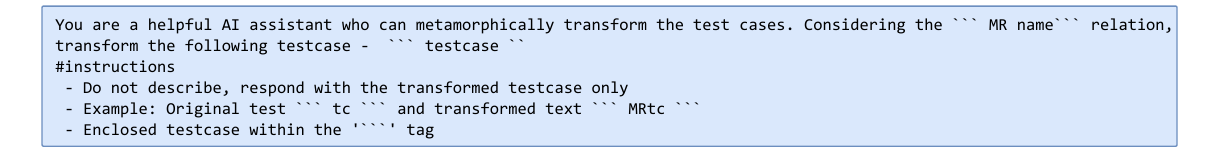}
    \caption{Prompt used to generate metamorphically transformed testcases}
    \label{fig:prompt_testmr}
\end{figure*}

\begin{figure}[!t]
    \centering
    \includegraphics[width=\linewidth]{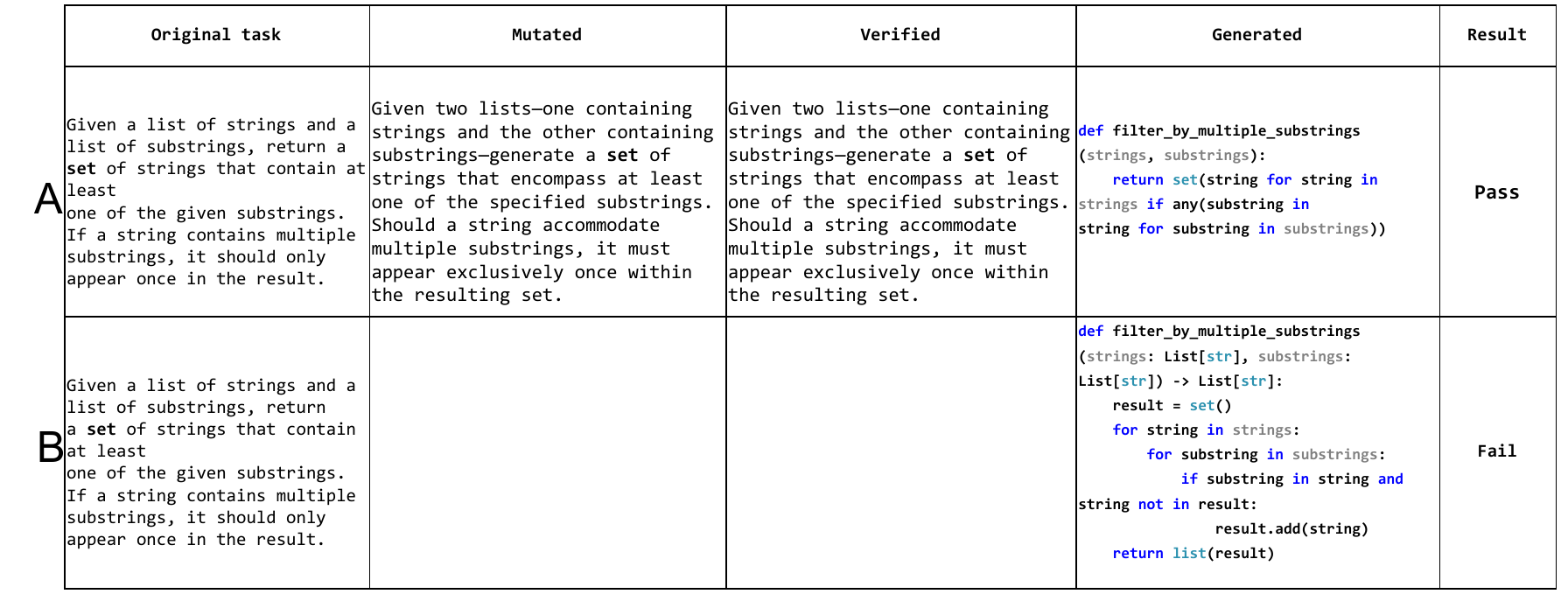}
    \caption{Example of the code generation process using LLM A) Following the CMA pipeline, and B) Without following the CMA mutation process}
    \label{fig:CMA_example_codegen}
\end{figure}

\begin{figure}[!t]
    \centering
    \includegraphics[width=\linewidth, trim = 0 0 0 0]{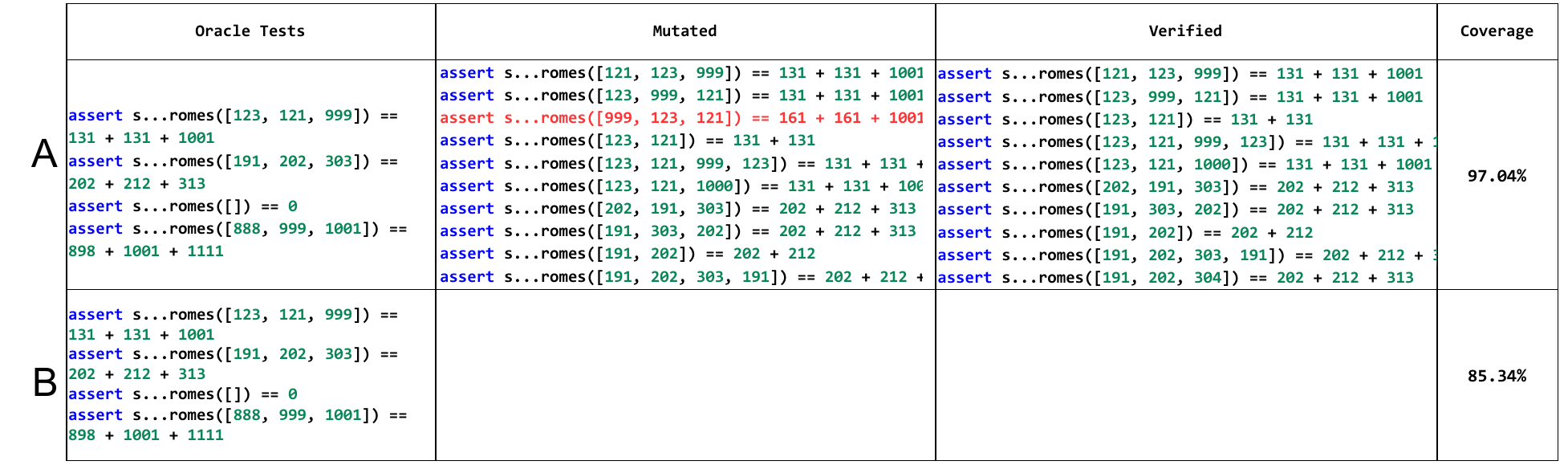}
    \caption{Example of the testcase generation process using LLM A) Following the CMA pipeline, and B) Without following the CMA mutation process}
    \label{fig:CMA_example_testgen}
\end{figure}

\subsection{Mutator}
The Mutator is the first component of the \textit{CMA} framework and is responsible for applying metamorphic relations to transform a given specification or test case into multiple semantically equivalent variants. It is designed to support transformation via either LLM-based or traditional machine learning operators; however, in the current implementation, we employ LLM-based transformations exclusively. These transformations include paraphrasing, logical inversion, procedural decomposition, and linguistic translation (see Section~\ref{sec:mr_selection}). Each metamorphic relation introduces a controlled semantic variation that preserves the underlying task intent while encouraging diverse reasoning trajectories from the LLM.

The Mutator operates across both the code generation and test case generation pipelines. For code generation and bug fixing tasks, it takes the original problem description $d$ and applies a set of metamorphic relations (MR1–MR4) to generate semantically equivalent variants $\{d_1, d_2, \dots, d_m\}$. The prompt template used for these transformations is shown in Figure~\ref{fig:prompt_mr}. For test case generation, the Mutator transforms the original test cases $t$ into a corresponding set of variants $\{t_1, t_2, \dots, t_k\}$ using relations MR5–MR9, as illustrated in Figure~\ref{fig:prompt_testmr}. Once the mutated descriptions or test cases are generated, they are forwarded to the Reviewer for semantic validation. Variants identified as invalid are returned to the Mutator for refinement, with a maximum of three iterations allowed.

To illustrate the Mutator's operation, ~\Cref{fig:CMA_example_codegen} shows an example of how a problem description is transformed using the Mutator. In the second column of row A, the original description is paraphrased to produce a semantically equivalent variant. In the original problem description, the requirement of returning a 'set' is less clear, but in the mutated description, it is explicitly mentioned to return a set of unique elements. This transformation helps the LLM better understand the requirement. Similarly, in the case of test case generation, ~\Cref{fig:CMA_example_testgen} illustrates how the original test cases are mutated with the Mutator. Each of the original test cases is transformed using different MRs to produce variants. For instance, the first test case is transformed using variable swapping, in the original testcase in input is an array with values \textsc{[123, 121, 999]}, from this, the first mutated testcase is generated with input \textsc{[121, 123, 999]}. With the mutation, the mutator generated 15 testcases from the original 4 testcases, given that the mutator filtered out duplicate testcases. As shown in the coverage column, though the original testcases achieved nearly 85\% coverage, the mutated testcases were able to improve test coverage to 97\%.  

\subsection{Reviewer}

The Reviewer ensures the semantic integrity of the variants produced by the Mutator. Its primary goal is to ensure that each transformed description or test case preserves the original functional intent without introducing ambiguity or altering the problem constraints. The Reviewer operates as a filtering and feedback mechanism that determines which variants are suitable for downstream generation and which require refinement.

For code generation and bug fixing tasks, the Reviewer analyzes the set of transformed problem descriptions $\{d_1, d_2, \dots, d_m\}$ using a BERT-based~\cite{reimers-2019-sentence-bert} semantic similarity model to quantitatively assess their alignment with the original specification $d_0$. Variants whose similarity scores fall below a defined threshold (0.8) are flagged as inconsistent and either discarded or returned to the Mutator for re-generation. It is obvious that, LLM can also be used as the reviewer, but with the intension to reduce the dependency over LLMs we choose the machine learning based model. 
This process ensures that only semantically faithful transformations are retained for subsequent prompting. The validated descriptions $\{d_1, d_2, \dots, d_n\}$ (where $n \leq m$) are then passed to the Generator for code synthesis. 

For test case generation, the Reviewer evaluates the MR-transformed test variants $\{t_1, t_2, \dots, t_n\}$ by executing them against the oracle implementation to verify behavioral equivalence. Test cases that produce inconsistent or invalid outputs are similarly filtered out or sent back to the Mutator for refinement. This dual validation process—semantic for specifications and behavioral for tests—maintains the fidelity of all metamorphic transformations within the framework. The validated test cases $\{t_1, t_2, \dots, t_k\}$ (where $k \leq n$) are then forwarded to the Evaluator for subsequent performance assessment.

Figure~\ref{fig:CMA_example_testgen} illustrates an example of the Reviewer filtering process,  an invalid mutated testcase (marked as red in the mutated list) is filtered out by the Reviewer before sending to the Evaluator. On the other hand, in Figure~\ref {fig:CMA_example_codegen}, the mutated description is semantically validated by the Reviewer before sending it to the Generator.

\begin{figure*}[t]
    \centering
    \includegraphics[width=\linewidth]{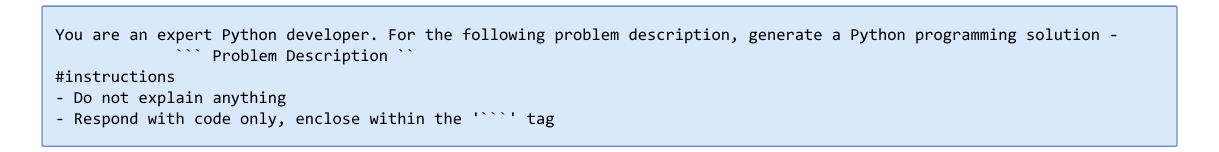}
    \caption{Prompt used to generate code from problem description}
    \label{fig:prompt_codegeneration}
\end{figure*}

\subsection{Generator}
The Generator is responsible for prompting the target LLM to produce code solutions or bug-fix patches from the validated variants provided by the Reviewer, as well as from the original problem description $d$ used as the baseline. It serves as the interface between the validated metamorphic transformations and the LLM’s content generation. Using structured prompt templates (e.g., Figure~\ref{fig:prompt_codegeneration}), the Generator ensures that all prompts remain consistent across tasks and model configurations.

For code generation and bug fixing tasks, the Generator takes the validated set of problem descriptions $\{d_1, d_2, \dots, d_n\}$ along with the original $d$ and, through the designated prompt template, independently queries the target LLM to produce corresponding code implementations or patches. One example is shown in Figure~\ref{fig:CMA_example_codegen}, where the Generator produces code solutions for the Verified Description (row A) as well as for the Original Description (row B). This approach allows for direct comparison between baseline and MR-guided generations. The generated solutions were subsequently forwarded to the Evaluator for execution and performance assessment. 

\subsection{Evaluator}

The Evaluator is the final component of the \textit{CMA} framework, responsible for assessing the correctness and behavioral consistency of the outputs generated by the LLM. It executes the generated code or patches and records their performance against benchmark-defined oracle tests to provide an objective measure of functional validity. The Evaluator also manages the comparison between baseline and MR-guided generations, ensuring that all variants are assessed under identical testing conditions.

For code generation and bug fixing tasks, the Evaluator executes both the baseline outputs (generated from $d$) and the MR-guided outputs (from $\{d_1, d_2, \dots, d_n\}$) using the provided oracle test suites. It records the number of passing tests and computes standard performance metrics such as functional correctness and success rate. Additionally, the Evaluator performs cross-variant checks to confirm that outputs derived from different metamorphic transformations produce equivalent results when executed on the same inputs.

For test case generation, the Evaluator measures the effectiveness of the validated MR-transformed test cases $\{t_1, t_2, \dots, t_k\}$ in terms of their contribution to test suite completeness. We used \texttt{coverage.py}\footnote{\url{https://coverage.readthedocs.io/en/7.11.0/}} to calculate the branch coverages to determine whether the new test cases exercise additional code regions beyond the original oracle suite.

Example results captured by the Evaluator are shown in Figures ~\ref{fig:CMA_example_codegen} and ~\ref{fig:CMA_example_testgen}. In the code generation example, the Evaluator records if the generated code pass or fail the oracle tests. In the test case generation example, it measures the coverage achieved by the given set of test cases. 

All evaluation results are systematically recorded and aggregated for subsequent analysis. We discussed the impact of metamorphic transformations on correctness, coverage, and robustness based the results conducted by the Evaluator.

Overall, the CMA transforms metamorphic relations from static testing artifacts into dynamic semantic operators that guide the entire LLM-driven software development process. Our experimental evaluations demonstrated that this agentic methodology leads to substantial improvements in correctness, bug resolution effectiveness, and test coverage across multiple LLM architectures and benchmark datasets.

\section{Metamorphic Relations}
\label{sec:mr_selection}

\begin{table*}[t]
\centering
\caption{Metamorphic Relations from the Literature}
\label{tab:metamorphic_relations}
\begin{tabular}{p{0.15\linewidth} p{0.30\linewidth} p{0.50\linewidth}}
\toprule
\textbf{Category} & \textbf{Metamorphic Relation} & \textbf{Description} \\
\midrule

\multirow{5}{*}{\centering\rotatebox{90}{\makecell[c]{Structural\\Input\\Transformations}}}

& Swapping \cite{liu2012new} & Swap the value positions in a function. (e.g. $f(a,b) = f(b,a))$  \\
& Permutation \cite{liu2012new} & Permuting the structure preserves relations. (e.g. $ f(X,a,b) = f(X', a', b')$)  \\
& Composite (swapping + permutation) \cite{liu2012new} & The swapping and permutation should preserve the relation (e.g. $f(X, a,b) = f(X', b', a')$)  \\
& Input-Output Equality \cite{chen2018metamorphic} & Relation between the input and output should be preserved after operations (if $ x -> x' $ and $y -> y'$, then $x = f(y)$ and $x'=f(y')$ ) \\
& Negation ~\cite{wu2025detecting} & Invert a comparative operator or logical predicate (e.g., \textit{max} $\leftrightarrow$ \textit{min}) \\

\midrule

\multirow{4}{*}{\centering\rotatebox{90}{\makecell[c]{Mathematical\\Operations}}}
& Left-Distributivity~\cite{Li2024MetamorphicRG} & Algebraic equivalence preserves output (e.g. $A\times(B+C)=A\times B+A\times C$). \\
& Incremental data \cite{xu2024mr} & Deterministic transformation such as shifting the date domain by $+1$ day should preserve valid behavior. \\
& Numeric/Boolean Expression Preservation ~\cite{Ayerdi2023GenMorphAG} & Applying algebraic identity transformations to numeric/boolean expressions should not alter functional correctness. \\
\midrule

\multirow{4}{*}{\centering\rotatebox{90}{\makecell[c]{Linguistic\\Variations}}}
& Semantic Similarity \cite{Wang2024MeTMaPMT} & Semantically similar phrases or sentences should lead to equivalent model interpretation and outputs. \\
& Semantic Difference \cite{Wang2024MeTMaPMT} & Semantically non-equivalent variants should not produce equivalent outputs. \\
& Chain-of-Though ~\cite{wu2025detecting} & Thinking of a step-by-step approach to solve the problem. \\
& Translation ~\cite{wu2025detecting} & Translating the input in a different language. \\
& Paraphrase ~\cite{Guo2024MORTARMM} & Rephrase a dialogue should remain stable semantically. \\
\midrule

\multirow{3}{*}{\centering\rotatebox{90}{\makecell[c]{Domain-\\Specific\\Adjustments}}}
& Domain-specific Subset ~\cite{li2024metamorphic} & Restrict inputs to a specific subset of the domain or scale them incrementally to observe stability and edge behavior. \\
& Equal Treatment ~\cite{Reddy2025MetamorphicTF} & Swapping demographic attributes shouldn't alter sentiment outcomes \\
& Sensitive Invariant ~\cite{Reddy2025MetamorphicTF}  & Model responses should be consistent across domains.
\\
\bottomrule

\end{tabular}
\end{table*}

The Mutator module in the proposed CMA agent relies on Metamorphic Relations (MRs) to systematically transform input problem descriptions into new variants. To identify the existing MRs suitable to CMA context, we conducted a comprehensive review of the prior research. Table \ref{tab:metamorphic_relations} summarizes the existing metamorphic relations from the literature. We identified a total of 12 unique metamorphic relations with corresponding applications. From the application perspective, the MRs can be broadly classified into four categories - 
\begin{itemize}
    \item Structural Input Transformations
    \item Mathematical Operations
    \item Linguistic Variations
    \item Domain-Specific Adjustments
\end{itemize}

To elaborate on these categories, the first category includes transformations that modify the structure of the input data while preserving its meaning. it includes operations such as swapping and permuting input elements. The MRs belongs to this category are particularly useful for testing the model's ability to handle variations. The second category encompasses transformations that apply based mathematical principles for data transformation. Rethar than sololy working on input domain, these MRs focus on the underlying mathematical relationships. The third category focuses on linguistic variations. These types of transformations are more relevant for natural language inputs and providing relational robustness in textual data. The final category of MRs are mostly domain-specific, these transformations are dependent on the context of the application domain. They ensure semantic equivalence or difference are preserved by the transformations.

After careful review of the collected MRs and their application in the corresponding literatures, we identified nine metamorphic relations (MRs) that are directly applicable to software engineering tasks, closely related to our CMA framework. From these pool of identified MRs, Negation (MR1), Translation (MR2), Redefining in steps (MR3), Paraphrasing (MR4) can be adjusted for linguistic transformations. We adopted these four MRs in task and issue description transformations. On the other hand, MRs which are mostly related input transformations and algebraic operations are more relevant for test case transformations. We opted Variable Swapping (MR5), Input Permutation (MR6), Algebraic/Distributive Transformation (MR7), Domain-specific Subset (MR8), and Incremental Data Transformation (MR9) for test case transformations.

\section{Experimental Design}
\label{experiment_design}

% This section describes the experimental design adopted to evaluate the effectiveness of \textit{Metamorphic Relations} (MRs) in improving Large Language Model (LLM) performance for code generation, bug fixing, and test case generation. We outline the datasets, model configurations, research questions, and experimental pipelines in this section. 
\subsection{Research questions} 
MRs have shown their ability in reducing hallucinations and increasing robustness with the LLM-generated outputs.   
Our goal is to identify the effectiveness of MRs within the LLM generation pipeline when they are applied to Software Engineering (SE). Among the SE tasks, LLMs are highly applied in code generation and validation steps. Hence, we designed two major research questions (RQs) for the experiment. 
\begin{enumerate}[label=\textbf{RQ\arabic{*}.}] 
    \item Can Metamorphic Relations help LLM in code generation tasks? 
    \item Can Metamorphic Relations help LLM to generate useful test cases? 
    \item How does individual MR perform in CMA for code generation tasks?
    % \item How does individual MR perform in CMA for test generation tasks?    
\end{enumerate} 

\subsection{Datasets}
We evaluate across multiple benchmarks that capture diverse aspects of LLM-based software engineering:

\begin{itemize}
    \item \textbf{HumanEval-Pro} \cite{yu2024humaneval}: an extended, higher-difficulty variant of the original HumanEval dataset, containing 164 programming problems with natural-language descriptions and oracle tests.
    \item \textbf{MBPP-Pro} \cite{yu2024humaneval}: an enhanced version of MBPP with 378 problems of varied complexity and clearer problem statements.
    % \item \textbf{SWE-Bench-Lite} \cite{jimenez2023swe}: a benchmark consisting of real-world software bugs derived from open-source repositories, used to evaluate LLM capabilities in automated bug fixing. The dataset contains 300 
\end{itemize}

Together, these datasets allow us to assess both synthetic and real-world scenarios, spanning code synthesis, repair, and testing.

\subsection{Model Selection}
\label{subsec:model_selection}

To ensure robustness and generality across architectures, we evaluate our framework using a diverse suite of state-of-the-art Large Language Models (LLMs) that differ in size, accessibility, and training data composition. The selection includes both proprietary and open-weight models, spanning multiple model families optimized for code understanding and synthesis.

Table~\ref{tab:model_config} summarizes the models, their parameter sizes, accessibility, and key configuration details. All models were evaluated in a zero-shot setting to isolate the impact of MR-guided refinement without additional fine-tuning or few-shot examples.

\begin{table}[ht]
\centering
\caption{Summary of evaluated Large Language Models (LLMs) used in experiments.}
\label{tab:model_config}
\begin{tabular}{p{0.22\linewidth}|p{0.15\linewidth}|p{0.60\linewidth}}
\textbf{Model} & \textbf{\# Parameters} & \textbf{Overview} \\
\hlineB{2}
\textbf{GPT-4o (2024-08-06)\cite{achiam2023gpt}} & Unknown & High-performance proprietary model accessed via OpenAI's official API. \\
\hline
\textbf{Mistral-Large (Instruct-2407)\cite{mistral_large}} & 123 Billion & Proprietary multilingual model with strong reasoning and coding capabilities, accessed via Mistral's official API. \\
\hline
\textbf{GPT-OSS \cite{agarwal2025gpt} } & 120 Billion & Open-weight GPT-style model specialized in code reasoning, deployed locally via Ollama. \\
\hline
\textbf{Qwen3-Coder  \cite{yang2025qwen3}} & 30 Billion & Open-weight code generation model, evaluated on a local GPU cluster using Ollama. \\
% \hline
% \textbf{CodeLlama (70B) \cite{roziere2023code}} & 70 Billion & Open-source LLaMA-based model fine-tuned for programming tasks, run locally via Ollama. \\
% \hline
% \textbf{DeepSeek-R1 \cite{guo2025deepseek}} & 70 Billion & Open-weight model optimized for code synthesis and repair, deployed locally on Ollama. \\
\hlineB{2}
\end{tabular}
\end{table}

Open-weight models (GPT-OSS, and Qwen3-Coder) were deployed locally using the \texttt{Ollama} AI model runner on a dedicated GPU server equipped with 160 GB of VRAM. Proprietary models (GPT-4o and Mistral Large) were accessed via their respective official APIs with identical prompt templates to maintain consistency across evaluations.

\subsection{Experimental Pipelines}
\label{subsec:pipelines}

\begin{figure*}[t]
    \centering
    \includegraphics[width=0.9\linewidth]{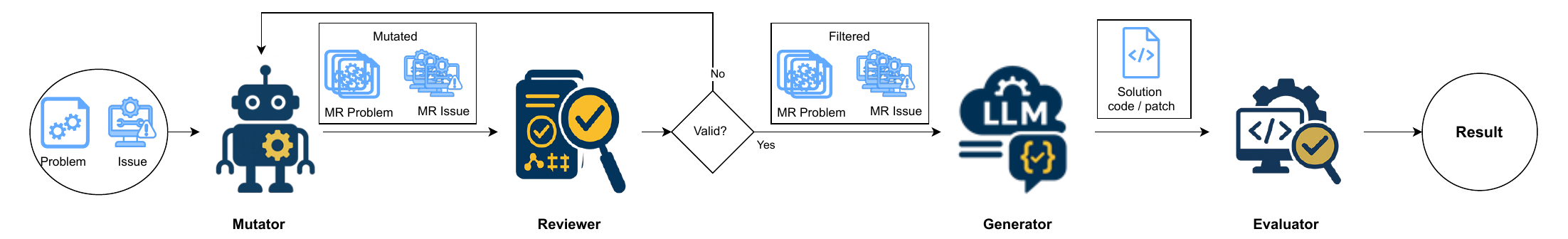}
    \caption{Code and bug fixing pipeline within \textsc{CMA}}
    \label{fig:mr_codegen_pipeline}
\end{figure*}

% \paragraph{MR-Guided Test Case Generation.}
% The second pipeline, shown in Figure~\ref{fig:mr_test_pipeline}, focuses on leveraging metamorphic relations to improve the diversity, coverage, and effectiveness of LLM-generated test suites. The process begins with either oracle test cases that accompany the benchmark problems. MRs are then applied to systematically transform these initial tests, generating new cases that target distinct but behaviorally related program states. 

% In parallel, the LLM is also prompted to generate test cases directly from the problem descriptions without MR guidance, serving as a baseline for comparison. All test sets—MR-derived and baseline—are executed against verified reference implementations. Their outputs are validated for correctness, redundancy, and contribution to coverage improvement. Coverage metrics include line, branch, and function coverage, while fault detection is assessed based on the number of unique defects exposed relative to the original oracle suite.

% The MR-guided process ensures that test generation is not only guided by structural or coverage heuristics but also grounded in semantic relationships between program behaviors. By comparing the effectiveness of MR-guided and baseline-generated test suites, we quantify how MRs enhance test completeness and the ability of LLMs to generate behaviorally meaningful tests.

\begin{figure*}[t]
    \centering
    \includegraphics[width=0.9\linewidth]{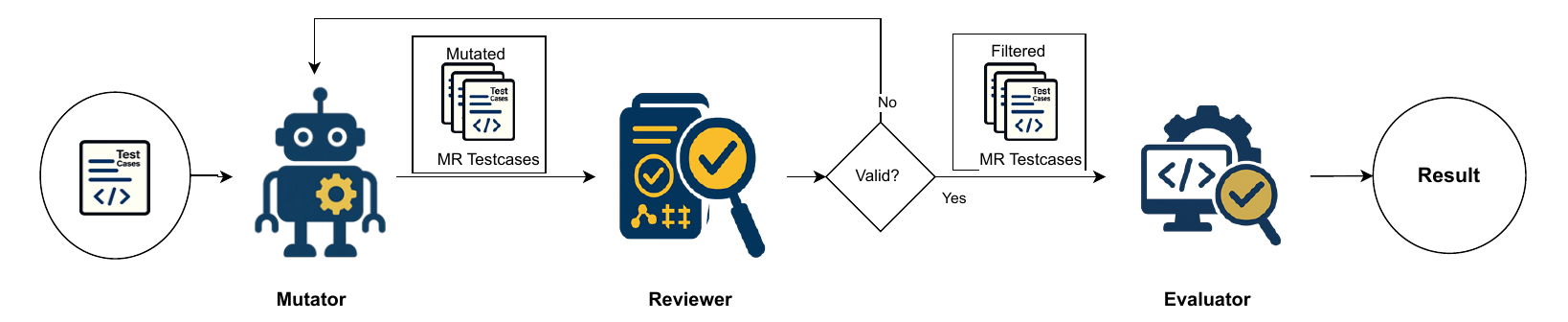}
    \caption{ Test case generation using  \textsc{CMA}}
    \label{fig:mr_test_pipeline}
\end{figure*}

The experiments were organized into two complementary pipelines corresponding to the major objectives of the study: (1) code generation and (2) test case generation. Both pipelines operationalize the \textit{CMA} framework to evaluate how metamorphic relations influence LLM performance across software engineering tasks.

As illustrated in Figure~\ref{fig:mr_codegen_pipeline}, it applies metamorphic transformations to problem descriptions or issue reports, generating multiple semantically equivalent variants for code synthesis and patch generation. The pipeline presented in  Figure~\ref{fig:mr_test_pipeline} applies metamorphic transformations to oracle test cases, producing logically consistent but structurally diverse variants for assessing test coverage and correctness. All generated outputs are executed against benchmark-provided oracle suites to ensure objective and reproducible evaluation.

These pipelines collectively provide a consistent setup for analyzing the impact of metamorphic relations on the accuracy, reliability, and completeness of LLM-generated software artifacts.

\section{Result and analysis}

% We analyze the role of MRs in generating code from problem specification with the help of LLMs.  

% \begin{enumerate}[label=\textbf{RQ1.\arabic{*}.}, leftmargin=35pt] 
%     \item Can CMA assist in generating more correct code from natural language specifications?
    
%     \item Can CMA help in fixing bugs based on issue descriptions?
    
% \end{enumerate} 

% \subsection{RQ1: Can CMA assist in generating more correct code from natural language specifications?}

\subsection{Effectiveness of CodeMetaAgent (CMA) on Guiding an LLM to Generate Correct Code (RQ1)}
\label{sec:RQ1_results}
% This research question is designed to investigate the impact of MRs for code generation tasks. 

\subsubsection{Motivation}

Metamorphic Relations (MRs) provide a systematic mechanism for refining task specifications and reducing linguistic ambiguity in code generation tasks. By transforming problem descriptions through semantically equivalent yet structurally distinct formulations, MRs can encourage Large Language Models (LLMs) to reason more consistently across problem representations. This is particularly valuable in code synthesis, where minor variations in phrasing can alter the model’s interpretation of the intended functionality, leading to incorrect or hallucinated implementations. By applying a suite of MRs-including negation, translation, redefining with steps, and paraphrasing-we aim to evaluate whether these transformations help LLMs better capture the underlying semantics of programming problems and thereby improve generation accuracy.

\subsubsection{Experiment}

For this experiment, we applied the set of MRs defined for code generation tasks (MR1–MR4) to each problem description from the \textit{HumanEval Pro} and \textit{MBPP Pro} benchmarks.  
Each original task was transformed into multiple semantically equivalent variants, which were then independently provided to the LLMs to generate corresponding code implementations.  
We evaluated three representative LLM models—\textit{GPT-OSS}, \textit{Qwen3-Coder}, \textit{Mistral} , and \textit{GPT-4o} - covering both open-weight and proprietary architectures.  
For each variant, we measured \textit{Pass@1} and \textit{Pass@5} scores by executing generated programs against oracle test suites.  
The integrated MR configuration combines all four MR transformations into a unified prompt-augmentation strategy.

\subsubsection{Result}

\begin{table}[!t]
\centering
\caption{Success rates (\%) of GPT-OSS, Qwen3-Coder, and Mistral models on Code generation for \textsc{HumanEval Pro} and \textsc{MBPP Pro} datasets }
\label{tab:codegeneration_stats}
\footnotesize
\begin{tabular}{l|l|ccc|ccc}
\toprule
\multirow{2}{*}{Model} &
\multirow{2}{*}{Dataset} &
\multicolumn{3}{c|}{Pass@1} &
\multicolumn{3}{c}{Pass@5} \\
\cmidrule(lr){3-5} \cmidrule(lr){6-8}
 &  & Base & CMA & Improvement 
 & Base & CMA & Improvement \\
\midrule
\multirow{2}{*}{GPT-OSS} 
 & HumanEval Pro & 60 & 69 & +9 & 61 & 70  & +9 \\
 & MBPP Pro      & 53 & 70 & +17 & 54 & 71 & +17 \\
\midrule
\multirow{2}{*}{Qwen3-Coder} 
 & HumanEval Pro & 76 & 83 & +7 & 78 & 84 & +6 \\
 & MBPP Pro      & 61 & 67 & +6 & 61 & 67 & +6 \\
\midrule
\multirow{2}{*}{Mistral} 
 & HumanEval Pro & 68 & 85 & +17 & 72 & 87 & +15 \\
 & MBPP Pro      & 60 & 66 & +6 & 62 & 68 & +6 \\
 \midrule
\multirow{2}{*}{GPT-4o} 
 & HumanEval Pro & 70 & 76 & +6 & 72 & 82 & +15 \\
 & MBPP Pro      & 65 & 67 & +2 & 66 & 68 & +2 \\

\bottomrule
\end{tabular}
\end{table}

Table~\ref{tab:codegeneration_stats} summarizes the results across both datasets. The column \textit{Base} represents the success rate of the solution generated using the original task description, while the column CMA represents the success rate achieved when applied all the MRs. Across the models, applying (\textit{MRs}) leads to a consistent and substantial improvement in both \textit{Pass@1} and \textit{Pass@5} metrics, indicating that MR-guided problem descriptions significantly enhance model comprehension and reliability.  

For \textit{GPT-OSS}, the CMA setup improved \textit{Pass@1} by +9\% on \textit{HumanEval Pro} and +17\% on \textit{MBPP Pro}. Note that we used the absolute improvements to represent the values, as these better reflect the practical utility gain. \textit{Qwen3-Coder} also exhibited stable gains, ranging from +6\% to +7\% across both datasets. 
\textit{Mistral} demonstrated one of the strongest improvements, especially on \textit{HumanEval Pro}, achieving +17\% (Pass@1) and +15\% (Pass@5). 
For \textit{GPT-4o}, improvements were smaller but still consistent, with +6\% on \textit{HumanEval Pro} and +2\% on \textit{MBPP Pro} (Pass@1).

Across models, the observed \textit{Pass@5} improvements follow similar patterns to \textit{Pass@1}. These results confirm that metamorphic mutations provide meaningful boosts over static task descriptions. Importantly, the consistent trend across models of different scales and types (open-weight vs. API-based) indicates that the performance gains are methodological and not tied to a specific model architecture or size.

\begin{figure*}[t]
\centering
\small
\begin{minipage}[t]{0.48\textwidth}
\begin{lstlisting}[caption={Original Task Description}, label={lst:task_org}]
Given a list of integers, determine the sum of the first 10 octagonal numbers for each integer in the list. If an integer is less than 1, return 0 for that integer.
\end{lstlisting}
% \caption*{}
\end{minipage}
\hfill
\begin{minipage}[t]{0.48\textwidth}
\begin{lstlisting}[caption={CMA Mutated Task Description}, label={lst:task_cma}]
1. Iterate through each integer in the list
2. Check if the integer is less than 1
3. If yes, return 0 for that integer
4. Otherwise, compute first 10 octagonal numbers
5. Sum the first 10 octagonal numbers
6. Return the aggregated total
\end{lstlisting}

\end{minipage}
\end{figure*}

To further illustrate the impact of CMA, we conducted a qualitative analysis of the generated code. Listing~\ref{lst:task_org} shows an example problem description from \textit{MBPP Pro} dataset. Which is transformed by CMA into the sequential steps as shown in Listing~\ref{lst:task_cma}. While Qwen3-Coder failed to generate a correct solution (Listing ~\ref{lst:code_org}) from the original description, but it successfully produced the correct implementation (Listing \ref{lst:code_cma}), when provided with the CMA-transformed prompt. With the original prompt, the model struggled to interpret the requirements of first 10 octagonal numbers and produced solution which does not adhere to the problem constraints. Listing~\ref{lst:code_cma} fails the test cases when input is greater than 10. However, when provided with the CMA-transformed prompt, it is evident that transformed prompt breaks down the problem into clear, sequential steps. Emphasizing the conditional logic and the specific requirement to calculate and sum the first 10 octagonal numbers CMA generated solution successfully passed all the oracle test cases. 

\begin{figure*}[t]
\centering
\small
\begin{minipage}[t]{0.48\textwidth}
\begin{lstlisting}[language=Python, caption={Code from Base description}, label={lst:code_org}]
def sum_of_octagonal_numbers(lst):
    def octagonal_number(n): return n*(3*n-2)
    result=[]
    for num in lst:
        if num<1: result.append(0)
        else:
            total=0
            for i in range(1,num+1): total+=octagonal_number(i)
            result.append(total)
    return result
\end{lstlisting}
% \caption*{(a) }
\end{minipage}
\hfill
\begin{minipage}[t]{0.48\textwidth}
\begin{lstlisting}[language=Python, caption={CMA Generated Code}, label={lst:code_cma}]
def sum_of_octagonal_numbers(lst):
    def octagonal_number(n): return n*(3*n-2)
    def sum_first_10_octagonal():
        total=0
        for i in range(1,11): total+=octagonal_number(i)
        return total
    sum_10=sum_first_10_octagonal()
    result=0
    for num in lst:
        if num>=1: result+=sum_10
    return result
\end{lstlisting}
% \caption*{(b) CMA Generated Code}
\end{minipage}
\end{figure*}

These results confirm that metamorphic mutations shows consistent improvements over the static description of tasks. Importantly, the consistent trend across architectures (ranging from 30B to 120B parameters) and access modalities (open-weight vs. proprietary APIs) indicates that the performance gains stem from the MR methodology rather than model-specific tuning.

% Qualitative inspection of model outputs reveals that MRs improve code clarity and help models to follow problem constraints better. For example, redefined or paraphrased prompts encourage step-wise reasoning and structured decomposition of the problem. Similarly, translation based reformulations reduce the reliance on lexical memorization.  
% As a result, LLMs generate code that is more logically consistent and less prone to hallucination or structural errors.

\addtocounter{o}{1}\lessons{\theo.Applying CMA to problem descriptions consistently increases code generation performance across models. The largest absolute gain reaches +17 points (GPT-OSS on \textsc{MBPP-Pro} and Mistral on \textsc{HumanaEval-Pro}).}

\addtocounter{o}{1}\lessons{\theo. Mistral achieves the highest absolute accuracy under CMA, reaching 87\% Pass@5 on \textsc{HumanEval-Pro}, showing that CMA strengthens even strong base models.}

\subsubsection{Case study : Bug fixing With MRs}

In the above experiment, we studied how mutation by MRs' helping code generation from task description. As code generation involved two types of scenarios - development from scratch and fixing the buggy codes. We also expended a case study to observe the MRs performance in generating codes (solution patches) to solve bugs. We employed CMA in the famous \textsc{SWE-Bench\_lite} ~\cite{yang2024swe} dataset to monitor the bug resolution rate from issue description. The dataset contains 300 issue description from popular GitHub repositories. For patch generation and evaluation we leveraged the Agentless ~\cite{agentless} pipeline with CMA. We observed original Agentless process with GPT-OSS(120B) model can generate 225 (75\%) patches and the bug resolution rate is 23\% (69 bugs in total). Alternatively when we applied CMA, the resolution rate improved to 30.3\% (90 bug). The number of generated patches is also raised to 233(78.6\%). These results demonstrate that MR-guided reformulation helps the model interpret bug reports more effectively. The contextual details MRs produce align better with the LLM’s representations. For example, sequential redefinition often transforms long, narrative-style issue descriptions into concise procedural instructions-highlighting reproduction steps and expected outcomes-which leads to more relevant code modifications. Similarly, paraphrasing and translation-based MRs help reduce lexical bias and force the model to generalize its reasoning beyond memorized phrasing.

\addtocounter{o}{1}\lessons{\theo. Applying metamorphic transformations to issue descriptions increases bug resolution success  7.3\%, confirming that MRs effectively improve LLM understanding of ambiguous issue reports.
}

\subsection{Effectiveness of CodeMetaAgent (CMA) on Guiding an LLM to Improve Test Coverages (RQ2)}
\label{sec:RQ2_results}

% The research question \textbf{RQ2.} evaluates whether MRs can enhance the test generation ability of LLM and contribute to software quality assurance. By comparing test cases generated using MRs with the test oracles, we aim to determine if MRs lead to generate an enhanced set of test cases and achieve better test coverage, particularly in uncovering edge cases. We analyze the test coverage improvement capabilities of MRs by answering the following sub-RQs: 

% \begin{enumerate}[label=\textbf{RQ2.\arabic{*}.}, leftmargin=35pt] 
%     \item Can the test cases generated using CMA improve testing performance?     
    
%     \item Can CMA improve testing performance compared to the tests generated from the problem description?
    
% \end{enumerate}

% \subsection{RQ2.1: Can the test cases generated using CMA improve testing performance?}

\subsubsection{Motivation}
Test case generation is fundamental to ensuring software correctness, reliability, and robustness. However, automatically generated tests from LLMs often suffer from redundancy, limited coverage, and a failure to expose subtle edge cases.  
Motivated by the useful adaption of metamorphic testing in software testing, MRs can be a guide to address these weaknesses by transforming input–output pairs into consistent but structurally varied test cases. 
By applying MRs, such as input swapping, permutation, distributive transformations, and incremental data modification—LLMs can generate diverse test cases that better capture program behavior across the input space.  
This research question investigates whether MR-based augmentation of test cases leads to measurable improvements in test coverage and validation performance, thereby improving the thoroughness of software quality assurance.

\subsubsection{Experiment}

We used the same benchmark dataset as \textbf{RQ1}, for this experiment. For each programming task, we started with the existing Oracle test suite and applied the metamorphic relations defined for test case generation (MR5–MR9).  
These transformations included input swapping, permutation, distributive expansion, domain-specific subsetting, and incremental data scaling.  
We employed different LLMs to generate additional test cases under both base and mutated configurations.  
Each generated test case was validated by executing it against the benchmark’s ground-truth implementation.  
Coverage metrics were computed using \texttt{coverage.py} , and improvements were measured relative to the baseline test suite.

To further study the impact MRs in LLMs' test generation, we asked LLMs to generate test-cases for the given problems. Each of the models was prompted to produce a complete test suite from the given problem without any MR augmentation. The same set of LLM models—\textit{GPT-4o}, \textit{Mistral-Large}, \textit{GPT-OSS (120B)}, and \textit{Qwen3-Coder (30B)} is also employed to this experiment. 

\begin{table}[!t]
    \centering
    \caption{Code coverage result (\%) MR-guided test generated by multiple LLMs compared against oracle tests}
    \label{tab:coverage_comparison}
    \footnotesize
    \begin{tabular}{lccccc}
        \toprule
        \textbf{Dataset} & \textbf{Oracle tests}  & \textbf {GPT-4o} & \textbf{Mistral} & \textbf {GPT-OSS} & \textbf{Qwen3-Coder} \\
        \midrule
        \textsc{MBPP Pro} & 99.36 & 99.81 &99.54 & 99.67 & 99.73 \\
        \textsc{HumanEval Pro} & 99.43 & 99.75 &99.52 & 99.74 & 99.76 \\
        \bottomrule
    \end{tabular}
\end{table}

\subsubsection{Results}
Table~\ref{tab:coverage_comparison} reports the overall coverage performance before and after MR-guided augmentation. Across all models and datasets, we observed consistent improvements, even in benchmarks that were already near-saturated in coverage. On \textsc{MBPP Pro}, the average coverage improved from 99.36\% in the base configuration to 99.81\% with GPT-4o, 99.54\% with Mistral, 99.67\% with GPT-OSS, and 99.73\% with Qwen3-Coder. Similarly, on \textsc{HumanEval Pro}, coverage increased from 99.43\% to 99.75\%, 99.52\%, 99.74\%, and 99.76\% respectively. While these numerical gains appear marginal due to the high baseline, a deeper analysis of coverage completeness reveals a more meaningful trend.

\begin{table}[!t]
\centering
\caption{Correctness rate (\%) of MR-generated test cases on \textsc{HumanEval Pro} and \textsc{MBPP Pro}.}
\label{tab:correctness_comparison}
\footnotesize
    \begin{tabular}{lcc}
        \toprule
        \textbf{Model} & \textbf{HumanEval Pro (\%)} & \textbf{MBPP Pro (\%)} \\
        \midrule
        GPT-4o           & 82.03& 88.15\\
        Mistral-Large    & 75.51& 76.55 \\
        GPT-OSS (120B)   & 78.23 & 87.35 \\
        Qwen3-Coder (30B)& 85.01 & 93.14 \\
        \bottomrule
    \end{tabular}
\end{table}

Additionally, we computed the correctness of mutated testcases. Table~\ref{tab:correctness_comparison} highlights  that, majority of MR-generated test cases were semantically valid. Qwen3-Coder demonstrated the highest correctness across both benchmarks, achieving 85.01\% on \textsc{HumanEval Pro} and 93.14\% on \textsc{MBPP Pro}, indicating strong robustness to metamorphic transformations. GPT-4o and GPT-OSS also produced highly reliable test cases, whereas Mistral, while exhibiting slightly lower correctness, still provided meaningful gains in coverage.

\begin{figure*}[t]
\centering
\small
\begin{minipage}[t]{0.85\textwidth}
\begin{lstlisting}[language=python, caption={Example Source code for test case evaluation}, label={lst:example_code}]
def next_smallest_palindrome(num):
    if all(d=='9' for d in str(num)): return num+2
    num=[int(d) for d in str(num)]
    n=len(num); mid=n//2; left_smaller=False
    i=mid-1; j=mid+1 if n%2 else mid
    while i>=0 and num[i]==num[j]: i-=1; j+=1
    if i<0 or num[i]<num[j]: left_smaller=True
    while i>=0: num[j]=num[i]; j+=1; i-=1
    if left_smaller:
        carry=1; i=mid-1
        if n%2:
            num[mid]+=carry; carry=num[mid]//10; num[mid]%=10; j=mid+1
        else: j=mid
        while i>=0:
            num[i]+=carry; carry=num[i]//10; num[i]%=10
            num[j]=num[i]; j+=1; i-=1
    return int("".join(map(str,num)))
def sum_of_next_smallest_palindromes(nums):
    if not nums: return 0
    return sum(next_smallest_palindrome(num) for num in nums)
\end{lstlisting}
\end{minipage}
\hfill
\begin{minipage}[t]{0.85\textwidth}
\begin{lstlisting}[language=python, caption={Oracle Tescases}, label={lst:oracle_tests}]
assert sum_of_next_smallest_palindromes([123, 121, 999]) == 131 + 131 + 1001
assert sum_of_next_smallest_palindromes([191, 202, 303]) == 202 + 212 + 313
assert sum_of_next_smallest_palindromes([]) == 0
assert sum_of_next_smallest_palindromes([888, 999, 1001]) == 898 + 1001 + 1111
\end{lstlisting}
\end{minipage}
\end{figure*}

\begin{minipage}[t]{0.85\textwidth}
\begin{lstlisting}[language=python, caption={CMA Tescases}, label={lst:cma_tests}]
assert sum_of_next_smallest_palindromes([121, 123, 999]) == 131 + 131 + 1001
assert sum_of_next_smallest_palindromes([123, 999, 121]) == 131 + 131 + 1001
assert sum_of_next_smallest_palindromes([123, 121]) == 131 + 131
assert sum_of_next_smallest_palindromes([123, 121, 999, 123]) == 131 + 131 + 1001 + 131
assert sum_of_next_smallest_palindromes([123, 121, 1000]) == 131 + 131 + 1001
assert sum_of_next_smallest_palindromes([202, 191, 303]) == 202 + 212 + 313
assert sum_of_next_smallest_palindromes([191, 303, 202]) == 202 + 212 + 313
assert sum_of_next_smallest_palindromes([191, 202]) == 202 + 212
assert sum_of_next_smallest_palindromes([191, 202, 303, 191]) == 202 + 212 + 313 + 202
assert sum_of_next_smallest_palindromes([191, 202, 304]) == 202 + 212 + 313
assert sum_of_next_smallest_palindromes([999, 888, 1001]) == 898 + 1001 + 1111
assert sum_of_next_smallest_palindromes([888, 1001, 999]) == 898 + 1001 + 1111
assert sum_of_next_smallest_palindromes([888, 999]) == 898 + 1001
assert sum_of_next_smallest_palindromes([888, 999, 1001, 888]) == 898 + 1001 + 1111 + 898
assert sum_of_next_smallest_palindromes([888, 999, 1002]) == 898 + 1001 + 1111

\end{lstlisting}
\end{minipage}

We present a concrete example to illustrate how MR-guided test generation increases branch coverage. Listing~\ref{lst:example_code} shows a program that computes the sum of the next smallest palindromes for a list of numbers. The oracle tests in Listing~\ref{lst:oracle_tests} cover only the common paths, achieving 85\% branch coverage. These tests execute normal flows, but they do not test several conditional outcomes such as even-odd digit handling or propagation inside the loop. In contrast, CMA-generated tests in Listing~\ref{lst:cma_tests} increase branch coverage to 97\%. Although the numerical values are closer, the MR-driven input variations force the program to execute both True and False outcomes of several internal decisions. This results in more diverse execution behavior and activates previously untested branches. The qualitative insight is that CMA expand behavioral exploration while preserving the original intent of the task.

These results collectively demonstrate that MR-guided test generation improves test suite effectiveness along two critical dimensions: \textit{completeness} (increased coverage and full coverage rates) and \textit{correctness} (high semantic validity of generated tests). This validates MRs as effective mutator for enhancing the robustness of LLM-generated test cases without requiring model fine-tuning or manual intervention.

\begin{table}[!t]
    \centering
    \caption{Code coverage (\%) on \textsc{MBPP Pro} and \textsc{HumanEval Pro} will LLM generated test cases from the task descriptions}
    \label{tab:coverage_with_LLM}
    \footnotesize
    \begin{tabular}{lcccc}
        \toprule
        \textbf{Dataset} & \textbf {GPT4} & \textbf{Mistral} & \textbf {GPT-OSS} & \textbf{Qwen3-Coder} \\
        \midrule
        \textsc{MBPP Pro}  & 99.56&98.14& 99.46& 99.52\\
        \textsc{HumanEval Pro} & 99.11&96.22& 95.89& 99.04\\
        \bottomrule
    \end{tabular}
\end{table}

In the Table~\ref{tab:coverage_with_LLM}, we are the LLMs' performance in generating test cases from the problem. Despite achieving relatively high coverages, this setup remain slightly below the results obtained through MR-guided augmentation (~\Cref{tab:coverage_comparison}). On \textsc{MBPP Pro}, GPT-4o achieved the highest baseline coverage (99.56\%), followed closely by Qwen3-Coder (99.52\%) and GPT-OSS (99.46\%), while Mistral lagged behind at 98.14\%. On \textsc{HumanEval Pro}, coverage dropped noticeably, particularly for GPT-OSS (95.89\%) and Mistral (96.22\%), suggesting comparative lacking in test case generation from problems.

\begin{table}[!t]
    \centering
    \caption{Correctness rate (\%) of MR-generated test cases on \textsc{HumanEval Pro} and \textsc{MBPP Pro}.}
    \label{tab:correctness_comparison_base}
    \footnotesize
    \begin{tabular}{lcc}
        \toprule
        \textbf{Model} & \textbf{HumanEval Pro (\%)} & \textbf{MBPP Pro (\%)} \\
        \midrule
        GPT-4o           & 78.23 & 82.18 \\
        Mistral-Large    & 72.11 & 75.38 \\
        GPT-OSS (120B)   & 72.35 & 82.55 \\
        Qwen3-Coder (30B)& 84.12 & 89.52 \\
        \bottomrule
    \end{tabular}
\end{table}

The correctness results in Table~\ref{tab:correctness_comparison_base} presents the correctness testcase generated following problem only. While Qwen3-Coder again demonstrated the highest correctness (84.12\% on \textsc{HumanEval Pro} and 89.52\% on \textsc{MBPP Pro}), these figures are consistently lower than those achieved under MR-guided augmentation (Table~\ref{tab:correctness_comparison}). Across all models, the average correctness improved by 3–6\% when MRs were applied, confirming that MR-based transformations reduce semantic inconsistencies and improve logical validity.

\begin{figure*}[t]
    \centering
    \includegraphics[width=0.9\textwidth]{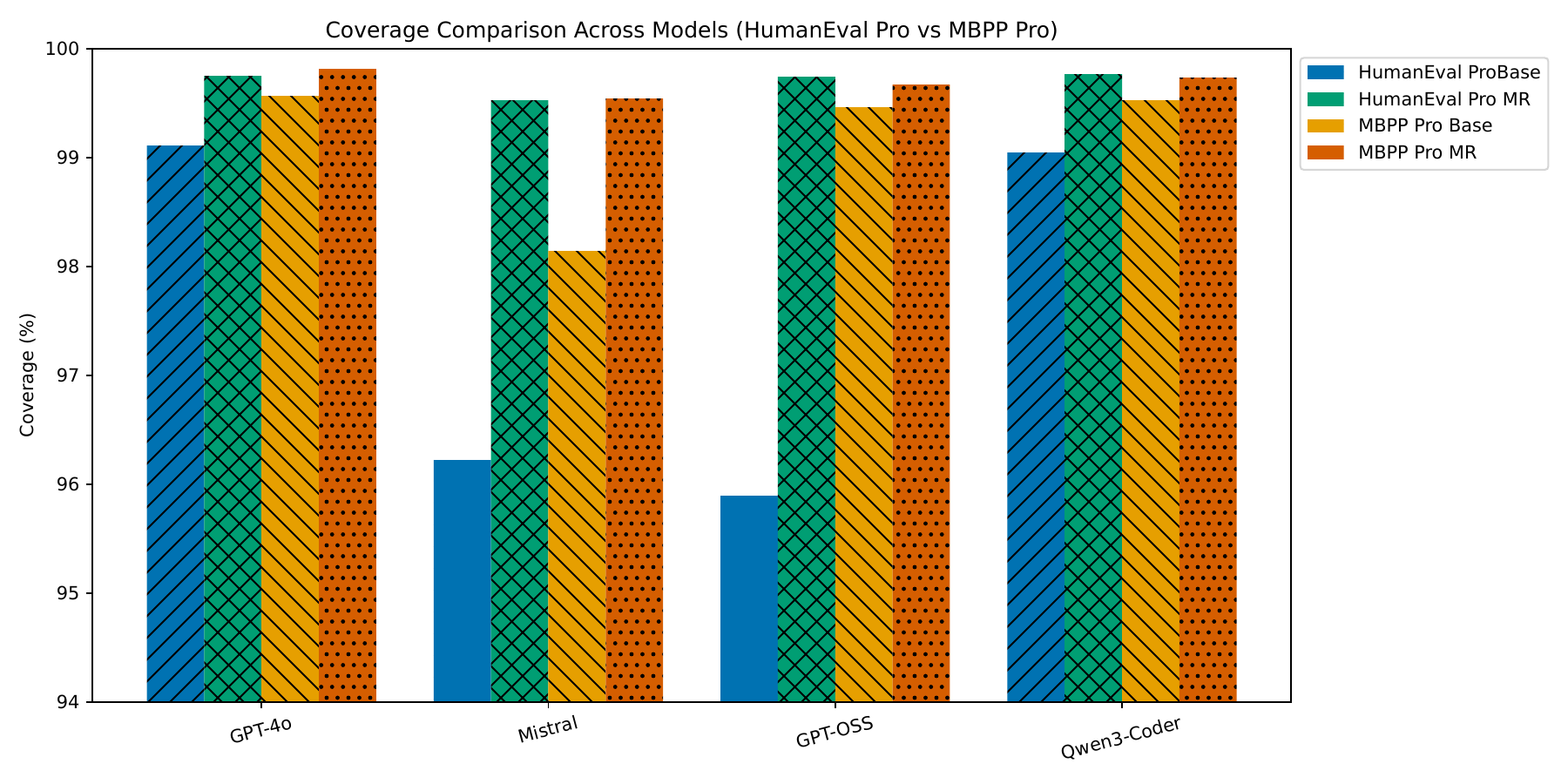}
    \caption{Grouped coverage comparison for test cases generated from problem descriptions vs MR-guided augmentation across \textsc{HumanEval Pro} and \textsc{MBPP Pro}. MR-guided tests consistently improve overall coverage across all models.}
    \label{fig:coverage_comparison}
\end{figure*}

\begin{figure*}[t]
    \centering
    \includegraphics[width=0.9\textwidth]{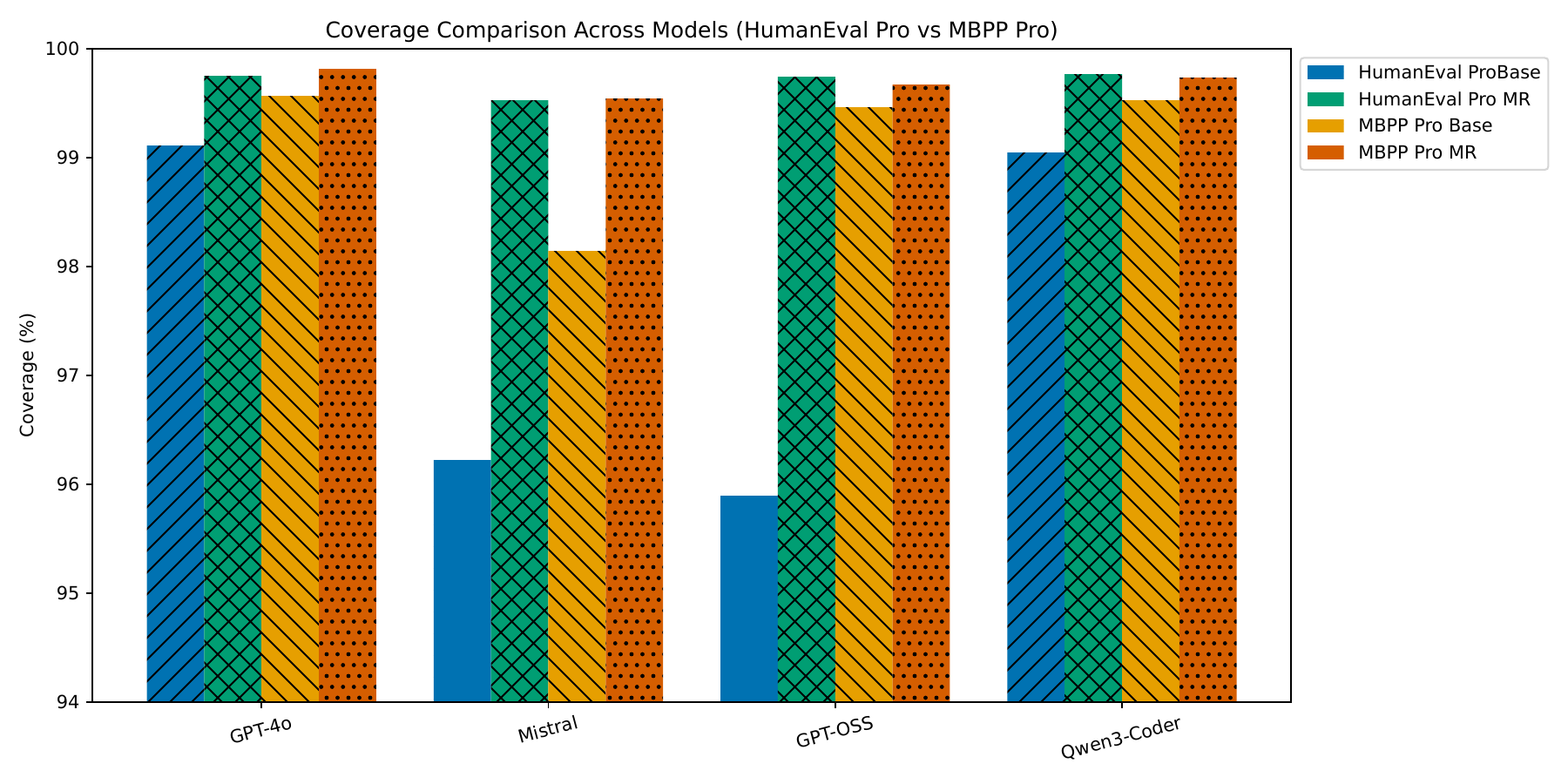}
    \caption{Grouped correctness comparison for test cases generated from problem descriptions vs MR-guided augmentation across \textsc{HumanEval Pro} and \textsc{MBPP Pro}. MR-guided tests exhibit higher semantic validity and reliability.}
    \label{fig:correctness_comparison}
\end{figure*}

Figures~\ref{fig:coverage_comparison} and~\ref{fig:correctness_comparison} provide visual comparisons of the LLMs' test case generation performance between problem-guided and MR-guided testcases. In the coverage comparison (~\Cref{fig:coverage_comparison}), MR-guided test suites consistently yield higher coverage across all models and both datasets, with the largest improvements observed for models with lower baseline performance such as Mistral (from 96.22\% to 99.52\% on \textsc{HumanEval Pro}, +3.30\%) and GPT-OSS (from 95.89\% to 99.74\%, +3.85\%). Even models with strong baseline performance, such as GPT-4o, exhibit measurable gains (from 99.11\% to 99.75\%, +0.64\%) on \textsc{HumanEval Pro}, and Qwen3-Coder improves from 99.04\% to 99.76\% (+0.72\%). This indicates that MR augmentation is particularly beneficial in closing residual coverage gaps, especially for models that struggle to infer diverse execution paths from natural language descriptions alone.

In the correctness comparison, the gap between MR-based and description-based test cases is even more pronounced. For example, GPT-4o improves from 78.23\% to 82.03\% (+3.80\%) on \textsc{HumanEval Pro}, and GPT-OSS increases from 72.35\% to 78.23\% (+5.88\%). Qwen3-Coder achieves the highest gains, improving from 84.12\% to 85.01\% on \textsc{HumanEval Pro} and from 89.52\% to 93.14\% on \textsc{MBPP Pro}, demonstrating that MRs enhance not only overall coverage but also the logical soundness and execution validity of generated test cases.

Although LLMs can generate reasonably complete test suites directly from the problem but MR-guided augmentation provides measurable benefits in both completeness and correctness. This improvement illustrates the ability of MRs to systematically explore variations of inputs, uncovering edge cases that problem–based test generation tends to overlook.

\addtocounter{o}{1}\lessons{\theo. MR-guided test augmentation consistently improves test completeness across all evaluated LLMs, with GPT-4o achieving the highest post-MR coverage of 99.81\% on \textsc{MBPP Pro}.}

\addtocounter{o}{1}\lessons{\theo. MR-generated test cases exhibit high semantic correctness, with Qwen3-Coder demonstrating the strongest reliability across both benchmarks, achieving up to 93.14\% correctness on \textsc{MBPP Pro}.}

\addtocounter{o}{1}\lessons{\theo. MR-guided transformations enhance both the coverage and correctness of solely LLM-generated test cases by 3–6\% on average, validating their role as an effective complement to description-based test generation.}

\subsection{How does individual MR perform in CMA for code generation tasks (RQ3)}

\subsubsection{Motivation}
CMA shows strong improvements when all MRs are applied together. However, it is not clear how each MR contributes individually. Different MRs modify the task in different ways, and some may help more than others. Understanding these differences can show which MRs are reliable and which may not be useful. Therefore, we study each MR separately to see how much benefit it provides in code generation tasks.

\subsubsection{Experiments}

\begin{figure}[h]
\centering
\subfloat[HumanEval-Pro (Pass@1)]{
    \includegraphics[width=0.46\linewidth]{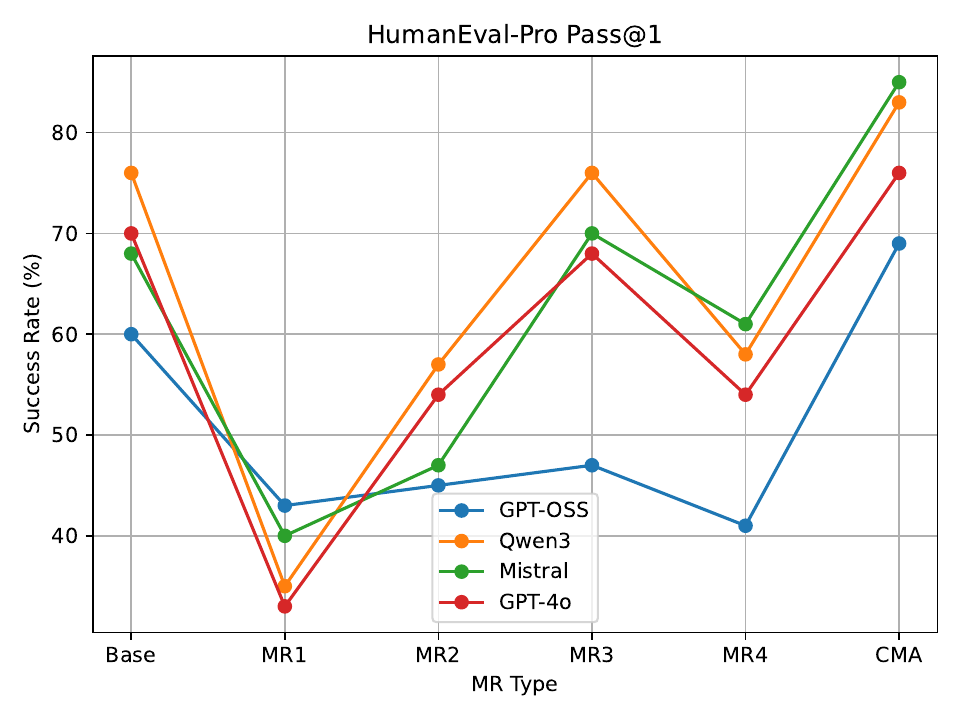}
}
\hfill
\subfloat[HumanEval-Pro (Pass@5)]{
    \includegraphics[width=0.46\linewidth]{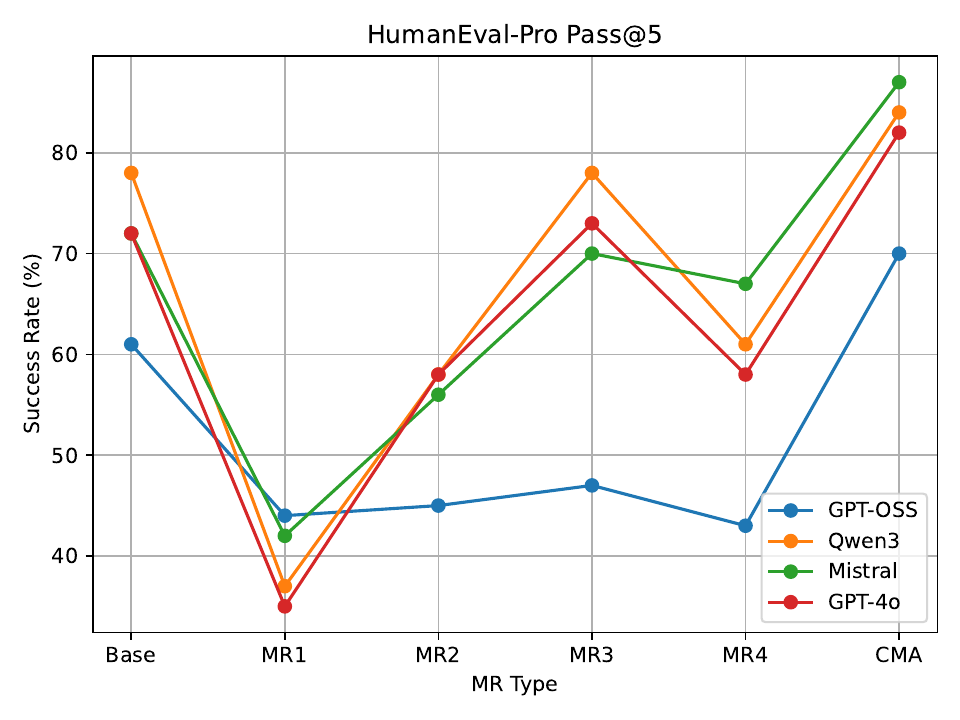}
}
\hfill
\subfloat[MBPP-Pro (Pass@1)]{
    \includegraphics[width=0.46\linewidth]{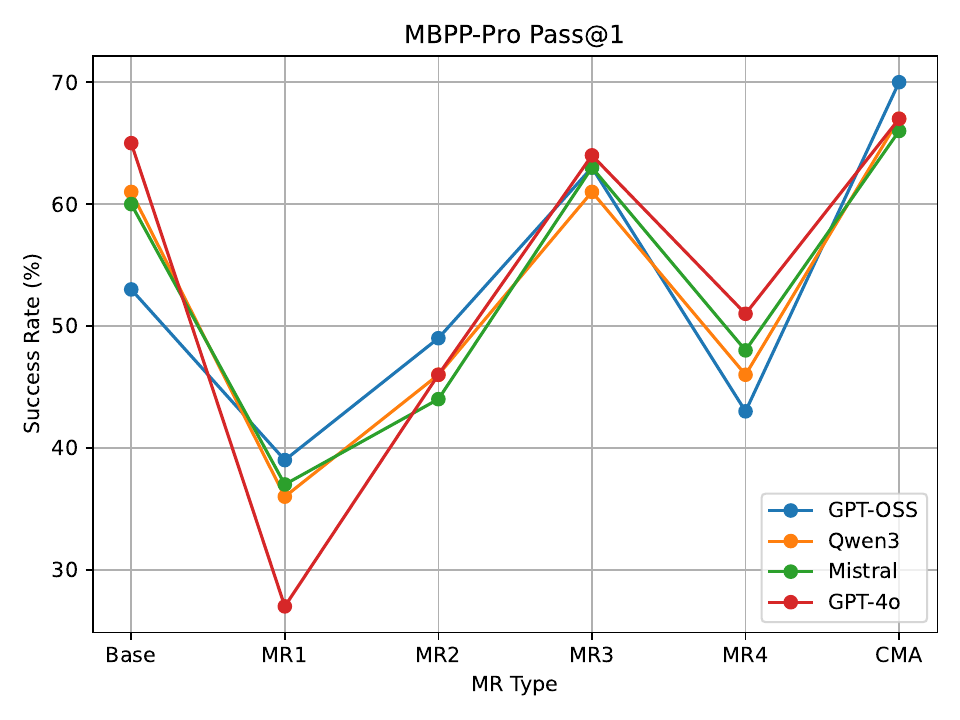}
}
\hfill
\subfloat[MBPP-Pro (Pass@5)]{
    \includegraphics[width=0.46\linewidth]{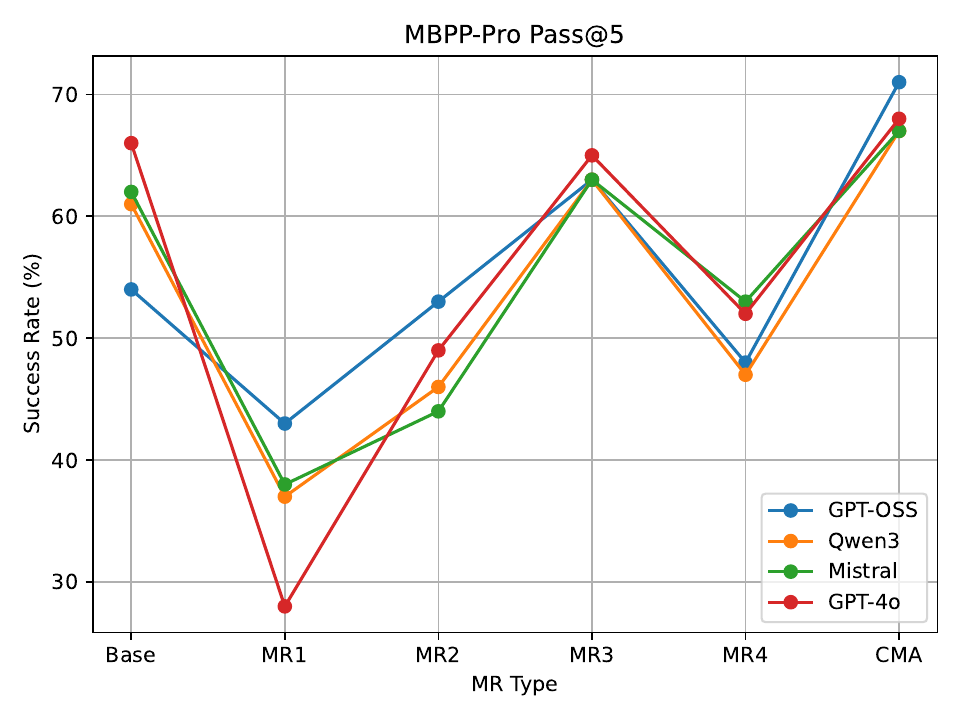}
}
\caption{Performance trends across MR variants for GPT-OSS, Qwen3, Mistral, and GPT-4o. (a) HumanEval-Pro Pass@1, (b) HumanEval-Pro Pass@5, (c) MBPP-Pro Pass@1, and (d) MBPP-Pro Pass@5 }
\label{fig:mrIndividualPerformance}
\end{figure}

For this study, we evaluate each MR independently. We apply one MR at a time to the problem description and generate code using that single transformed variant. This helps isolate the direct effect of each MR. We test four LLMs (GPT-OSS, Qwen3-Coder, Mistral, and GPT-4o) on \textsc{HumanEval-Pro} and \textsc{MBPP-Pro}. We report the results using Pass@1 and Pass@5 metrics. The CMA line in Figure~\ref{fig:mrIndividualPerformance} represents the setup where all MRs are applied together.

% \subsection{Experiments}

% \setlength{\tabcolsep}{3pt}
% \begin{table}[h]
% \centering
% \caption{Individual MR performance comparison}
% \label{table:mrIndividualPerformance}
% \footnotesize
% \begin{tabular}{l|cccc|cccc|cccc|cccc}
% \hline
%  & \multicolumn{8}{c|}{\textbf{Pass@1}} & \multicolumn{8}{c}{\textbf{Pass@5}} \\
% \textbf{MR Type} & \multicolumn{4}{c|}{HumanEval-Pro} & \multicolumn{4}{c|}{MBPP-Pro} & \multicolumn{4}{c|}{HumanEval-Pro} & \multicolumn{4}{c}{MBPP-Pro} \\
%  & GPT-OSS & Qwen3 & Mistral & GPT-4o & GPT-OSS & Qwen3 & Mistral & GPT-4o & GPT-OSS & Qwen3 & Mistral & GPT-4o & GPT-OSS & Qwen3 & Mistral & GPT-4o \\
% \hline
% Base   & 60 & 76 & 68 & 70 & 53 & 61 & 60 & 65 & 61 & 78 & 72 & 72 & 54 & 61 & 62 & 66 \\
% MR1    & 43 & 35 & 40 & 33 & 39 & 36 & 37 & 27 & 44 & 37 & 42 & 35 & 43 & 37 & 38 & 28 \\
% MR2    & 45 & 57 & 47 & 54 & 49 & 46 & 44  & 46 & 45 & 58 & 56 & 58 & 53 & 46 & 44  & 49 \\
% MR3    & 47 & 76 & 70 & 68 & 63 & 61 & 63 & 64 & 47 & 78 & 70 & 73 & 63 & 63 & 63 & 65 \\
% MR4    & 41 & 58 & 61 & 54 & 43 & 46 & 48 & 51 & 43 & 61 & 67 & 58 & 48 & 47 & 53 & 52 \\
% CMA & 69 & 83 & 85 & 76 & 70 & 67 & 66 & 67 & 70 & 84 & 87 & 82 & 71 & 67 & 67 & 68 \\
% \hline
% \end{tabular}
% \end{table}

% We also analyzed the performance of individual MRs in the code generation tasks. The figure ~\ref{fig:mrIndividualPerformance}  represents the comparative results for each of the MRs. We observed that, the success rate of Negation (MR1) and  translation (MR2) generated code remains close to the code generated from the original description, while in some cases MR1 fall behind. The performance drop with MR1, indicated that the double negation of the statements sometimes alter the underlying meaning of the statement. The step by step refinement of the task (MR3) gained highest success in every combination, as it allowed LLM to reason the problem more deeply and generate a clearer programming flow for the given problem. Paraphrasing (MR4) also consistently improved performance over other setups, excepts the (MR3). The trend remain similar for all the experimented LLMs.      
% Another observation is that the pass@5 success rate is highly analogous to the pass@1 performance. We found that across multiple runs on the same prompt the LLM often produce same code repeatedly with only a few exceptions.

% The performance contribution of each individual MR in the code generation tasks.

\subsubsection{Results}
Figure~\ref{fig:mrIndividualPerformance} illustrates the comparative performance of each MR variant across models and datasets. Negation (MR1) and Translation (MR2) generally remain close to the baseline, although MR1 fall below the original description due to meaning distortion from polarity inversion. For example, Qwen3 drops to 35\% Pass@1 under MR1 on \textsc{HumanEval-Pro}, while CMA yields 83\%. MR2 shows modest gains, such as GPT-OSS reaching 45\% Pass@1 compared to its 60\% baseline.

In contrast, MR3 consistently demonstrates the highest positive impact. Step-wise decomposition strengthens reasoning, enabling models to generate more structured and logically aligned code. MR3 reaches 78\% Pass@5 on \textsc{HumanEval-Pro} with Qwen3, approaching the CMA upper bound of 84\%. Paraphrasing (MR4) also improves performance across all models, though never surpassing MR3. For instance, GPT-4o reaches 54\% Pass@1 under MR4 but increases further to 68\% under MR3.

It is evident that MR3's structured reformulation most effectively guides LLMs toward correct solutions, while MR1 and MR2 offer limited benefits. Morespecifically, MR1's negation can inadvertently alter task semantics, leading to confusion, and it is not sufficiently helping the model to reason about the problem correctly.MR2's translation may introduce linguistic artifacts that hinder comprehension. CMA mostly benefited from MR3+MR4 combinations, which together enhance clarity and reasoning depth. Their combined effect likely drives the majority of CMA's overall performance gains.

% These patterns remain stable across GPT-OSS, Qwen3, Mistral, and GPT-4o. Pass@5 trends closely match Pass@1, showing minimal additional diversity from sampling. This indicates that performance gains originate primarily from structural reformulation driven by MR-induced reasoning changes, rather than from multiple sampling attempts. 

% \begin{lstlisting} [caption]
% Base: Given a list of side lengths of squares, write a function that returns the total perimeter of all the squares

% Negation: Do not fail to write a function that returns the total perimeter of all the squares given a list of side lengths of squares.

% Steps-wise refinement: 
% 1. Receive a list of side lengths as input parameter
% 2. Initialize a variable to store the total perimeter sum
% 3. Iterate through each side length in the input list
% 4. For each side length, calculate the perimeter of the corresponding square (4 times the side length)
% 5. Add each square's perimeter to the total sum
% 6. Return the final total perimeter sum

% Rewrite: 
% Given a collection of side length measurements corresponding to multiple squares, develop a function that computes and returns the aggregate perimeter encompassing all specified squares.
      
% \end{lstlisting}

\addtocounter{o}{1}\lessons{\theo.  MR3 shows the highest success rate among the MRs, with highest success ratio 78\% when applied to Qwen3-coder for Humaneval-Pro}

% \subsection{How does individual MR perform in CMA for test generation}

% \subsubsection{Motivation}

% \subsection{Experiments}

% \section{Result}

\section{Discussion}

\subsection{Token analysis}

LLM-based models rely on tokenized text representations. The number of tokens directly affects computation and cost. In the \textit{CMA}, the Mutator and Generator modules trigger LLM requests. The Mutator module uses about 205 input and 287 output tokens per problem. The Generator module consumes around 258 input and 328 output tokens. Notably, the Mutator’s token cost occurs only once across all models. In contrast, code generation from the Oracle description requires just 106 input and 91 output tokens. Hence, \textit{CMA} demands roughly four times more input and five times more output tokens. These results confirm that agent-assisted processing enhances reasoning power but increases token utilization and inference cost.

% \begin{table}[!t]
% \centering
% \caption{Token and cost analysis per MetacodeAgent stage. Costs computed per request using public API rates\textsuperscript{*}. $\Delta$ columns report percentage increase relative to the baseline.}
% \label{tab:mca_token_costs}
% \footnotesize
% \begin{tabular}{lrrrrrrr}
% \toprule
% \textbf{Stage} & \textbf{Input} & $\boldsymbol{\Delta}$\textbf{In} & \textbf{Output} & $\boldsymbol{\Delta}$\textbf{Out} & \textbf{GPT-4o Cost} & \textbf{Mistral Cost} & $\boldsymbol{\Delta}$\textbf{Cost} \\
% \midrule
% Base (single prompt) & 106 & 0.0\% & 91  & 0.0\% & \$0.001895 & \$0.000758 & 0.0\% \\
% Mutator               & 205 & 93.4\% & 287 & 215.4\% & \$0.005330 & \$0.002132 & 181.3\% \\
% Generator             & 258 & 143.4\% & 328 & 260.4\% & \$0.006210 & \$0.002484 & 227.7\% \\
% \bottomrule
% \end{tabular}

% \vspace{3pt}
% \raggedright \footnotesize \textsuperscript{*}Assumes API prices per 1M tokens: GPT-4o (input \$5, output \$15); Mistral Large (input \$2, output \$6). Cost = (input\_tokens/1e6)\,$\times$\,price\_in + (output\_tokens/1e6)\,$\times$\,price\_out. 
% \end{table}

% base input token: 106 (avg)
% output tokens: 91
% MCA:
%     Mutator Input tokens: 205
%     Mutator Output token: 287 
%     Generator input token: 258
%     Generator output token: 328

\subsection{Making CMA Sustainable}

It is important to address the increasing token consumption in the \textit{CMA} to ensure the viability. Though the performance gains are significant, the elevated token usage may pose challenges for scalability and cost-effectiveness. 
To reduce the reliance on expensive LLM calls, one approach can be to optimize the Mutator module. It can incorporate a symbolic or rule-based algorithm to mutate the input. Alternatively, it can be designed with a low-cost or small LLM-based approach to make the framework more cost-effective.

Another approach can be adaptive token budget management, applying MRs selectively based on utility estimates. For example, stop mutation rounds when sufficient achievements are already gained or incremental gain diminishes. 

Another alternative can be using a multilevel model execution strategy. In the first level, low-cost or local LLM models can be used for mutation and cache the results for later use. The next level would be designed with more powerful or premium LLMs for the final generation or complex scenarios where the first level fails. In addition, keeping a small set of high-impact MRs can prevent a higher number of variant generations and improve efficiency.

% Resource Usage and Cost

% Latency

% Memory Management Techniques

\subsection{Recommendations}

% \todo{
% update observation numbers
% }

\begin{table}[!t]
    \centering
\caption{Mapping of Recommendations to Research Observations ( Observations marked as O)}
\label{tab: recommendations}
\begin{tabular}{p{0.80\linewidth} p{0.17\linewidth}}
\toprule
\textbf{Recommendation} & \textbf{Supported By} \\
\midrule
\rowcolor{lightgray}
\textbf{R1:} Pre-process LLM inputs using MRs & O1, O2, O3 \\
\textbf{R2:} Adopt MR-guided test case mutation & o4, O5, O6 \\

\rowcolor{lightgray}
\textbf{R3:} Incorporate MR as a core prompting strategy & O1, O3, O4 \\
\textbf{R4:} Investigate the integration of MRs with other prompt strategies & O1, O6 \\
\rowcolor{lightgray}
\textbf{R5:} Explore the application of MRs to other Software Engineering tasks & O3, O4 \\

\rowcolor{lightgray}
\textbf{R6:} Selecting the correct set of MRs is a key & O7 \\

\textbf{R7:} Develop a standardized benchmark for evaluating MRs in Software Engineering & All \\
\rowcolor{lightgray}
\textbf{R8:} Generate a rule-based mutation system to reduce LLM API calls & All \\
\rowcolor{lightgray}
\textbf{R9:} Agent helps to Automate MR selection and improve LLMs' efficiency & All \\
\bottomrule
\end{tabular}
\end{table}

In Table~\ref {tab: recommendations}, we offer several recommendations to improve the use of MRs based on our study. Each recommendation is mapped to one or more observations from the experiments. Details of the recommendations are discussed below -  

\begin{enumerate}[label=\textbf{R\arabic{*}.}] 
    \item \textbf{Pre-process LLM inputs using MRs}

    As we discussed in the Section~\ref{sec:RQ1_results},  the mutated descriptions using MRs can unveil different facets of the problem statement, which help LLMs consistently generate more accurate solutions. Therefore, we recommend pre-processing the input prompts using MRs to enhance the LLM's comprehension before generating solutions.

    \item \textbf{Adopt MR-guided test case mutation}
    
    Based on the findings from Section~\ref{sec:RQ2_results}, we observed that mutating test cases using MRs can effectively identify edge cases and improve the robustness of the generated code. We recommend incorporating MR-guided test case mutation strategies to enhance the evaluation and validation of LLM-generated solutions.

    % \item \textbf{Select LLMs Based on Task}
    
    \item \textbf{Incorporate MR as a core prompting strategy} 
    
    While R1 and R2 uses MRs as pre-processing, this recommendation suggests embedding MRs directly into the prompt structure. Following the observations O1, O3, and O5, we suggest that model that prompted with  MR-augmented data outperform better. Therefore, we recommend integrating MRs as a fundamental component of the prompting strategy. 

    \item \textbf{Investigate the integration of MRs with other prompt strategies}
    Observations O1 and O7 shows that MRs are powerful assintant to LLMs. Hence, we recommend exploring the synergy between MRs and other prompting techniques, such as self-consistency, Retrieval-Augmented Generation (RAG), or few-shot prompting, to further enhance LLM performance in software engineering tasks.

    \item \textbf{Explore the application MRs to other Software Engineering tasks}
    Building on the success observed in code generation, bug fixing, and test mutation, we recommend investigating the applicability of MRs to other software engineering tasks, such as code refactoring or documentation generation to further extend the benefits of MR-guided approaches.

    \item \textbf{Selecting the correct set of MRs is a key}
    Our findings show that not all MRs contribute equally to performance. Some MRs (such as step decomposition and paraphrasing) consistently act as performance boosters, while others (like negation) can degrade results depending on the semantic structure of the task. We therefore recommend that practitioners selectively activate only high-impact MRs rather than applying all MRs uniformly. This selective MR activation can prevents semantic distortion, reduces unnecessary computation, and ensures more reliable gains in real deployment settings.

    \item \textbf{Develop a standardized benchmark for evaluating MRs in Software Engineering}

    Our observations highlight MRs' effectiveness across various Software Engineering tasks. Also from the literature, there is a growing interest in MRs for Software Engineering tasks. However, there is a lack of standardized benchmarks to evaluate their impact comprehensively. We recommend developing a benchmark that help comparing different MR-based approaches and their effectiveness across diverse tasks. A standardized benchmark will prevent cherry-picking results and promote fair comparisons.
    
    \item \textbf{Generate a rule-based mutation system to reduce LLM API calls}
    
    The methodology employed in \textit{CMA} relies heavily on LLM calls. The Mutator and Generator modules both invoke LLMs, leading to increased token usage and associated costs. To facilitate by LLM generation Generator is bounded to use LLMs. However, the Mutator module can be redesigned to use rule-based or symbolic mutation techniques. A rule-based mutation system can be an effective alternative to LLM-based mutation. Although, there are existing text transformation tools , they are not specifically designed for MR specific mutations. Therefore, we recommend developing a dedicated rule-based mutation system that can perform MR-guided transformations accurately.  

    \item \textbf{Agent helps to Automate MR selection and improve LLMs' efficiency} 
    We found that selecting the right set of MRs is crucial for maximizing performance gains. However, agenets can be designed to automatically select the most appropriate MRs based on the task context and model capabilities. Agents can help reducing unnecessary computations by avoiding low-impact MRs as well as optimize computation by caching and reusing mutated variants. Therefore, we recommend developing an agent-based system that can intelligently select and apply MRs to enhance LLM efficiency and effectiveness in Software Engineering tasks.

\end{enumerate}

\subsection{Limitations} 

While \textit{CMA} demonstrates consistent improvements across multiple benchmarks and models, several limitations must be acknowledged. The current implementation employs only LLM-based mutation operators within the Mutator module, which may introduce linguistic or domain biases inherited from the underlying models. Integration of symbolic and ML-based transformations could help reduce the dependencies on LLM models and improve cost efficiency. In addition, the Evaluator primarily focuses on functional correctness and code coverage, leaving other quality dimensions such as efficiency, maintainability, and semantic equivalence unexplored. Experiments were conducted using Python-based benchmarks (\textsc{HumanEval Pro}, \textsc{MBPP Pro}, and \textsc{SWE-Bench-Lite}), which may limit generalization to other programming languages or industrial-scale systems. Furthermore, proprietary APIs (e.g., GPT-4o, Mistral-Large) and their evolving configurations impose partial reproducibility constraints beyond the authors’ control.

\subsection{Threats to Validity}

The validity of the study can be threatened by several factors. First, the validity of the results may depend on the quality of the benchmark tasks and the MR transformations applied. The effectiveness of MR-guided refinement is influenced by the quality of the prompts, the reasoning capacity of the underlying LLM, and the representativeness of the task examples used for mutation. Second, for open-weight models such as GPT-OSS and Qwen3-Coder, the computational hardware used for local inference may have affected latency or generation quality. Third, implementation choices—such as prompt structure, evaluation scripts, or Reviewer thresholds—may introduce unintended bias. Furthermore, due to the non-deterministic generative nature of LLMs, reproducing identical outputs across runs is inherently challenging, particularly when using stochastic decoding or API-based models. Despite these potential threats, we minimized their impact by employing deterministic decoding settings, consistent prompt templates, and standardized evaluation pipelines across all experiments.

\section{Conclusion}

We introduced \textit{CodeMetaAgent(CMA)}, an LLM-agent that enhances the reasoning and reliability of large language models in software engineering tasks through the integration of metamorphic relations (MRs). By coordinating four core modules—Mutator, Reviewer, Generator, and Evaluator—the framework enables LLMs to reason over semantically equivalent task variants, refine specifications, generate solutions, and validate outputs systematically. Experiments on \textsc{HumanEval Pro}, \textsc{MBPP Pro}, and \textsc{SWE-Bench} benchmarks demonstrate substantial performance gains, achieving up to 17\% improvement in code generation accuracy. The results establish CMA as the first agentic framework to operationalize MRs within LLM reasoning. While deeper observation to individual MRs' performance emphasize on selecting MR is a key for gaining better achievements. For the code generation oriented tasks transformations like negation and translation didn't achieve significant improvements, whereas step-wise decomposition and paraphrasing substantially boosted performance across all evaluated models. On the other hand, for the test generation tasks, MR's enhances test cases to capture edge cases and improve code coverage even with problem having very limited contexts. 
  Future work will extend this foundation through multi-agent collaboration, integration of symbolic reasoning, and adaptive MR selection, further advancing the reliability, autonomy, and trustworthiness of LLM-driven software engineering.

\bibliographystyle{ACM-Reference-Format}
\bibliography{main}

@article{yu2024humaneval,
  title={HumanEval Pro and MBPP Pro: Evaluating Large Language Models on Self-invoking Code Generation},
  author={Yu, Zhaojian and Zhao, Yilun and Cohan, Arman and Zhang, Xiao-Ping},
  journal={arXiv preprint arXiv:2412.21199},
  year={2024}
}

@article{zhou2018metamorphic,
  title={Metamorphic relations for enhancing system understanding and use},
  author={Zhou, Zhi Quan and Sun, Liqun and Chen, Tsong Yueh and Towey, Dave},
  journal={IEEE Transactions on Software Engineering},
  volume={46},
  number={10},
  pages={1120--1154},
  year={2018},
  publisher={IEEE}
}

@inproceedings{mayer2006empirical,
  title={An empirical study on the selection of good metamorphic relations},
  author={Mayer, Johannes and Guderlei, Ralph},
  booktitle={30th Annual International Computer Software and Applications Conference (COMPSAC'06)},
  volume={1},
  pages={475--484},
  year={2006},
  organization={IEEE}
}

@article{wu2025detecting,
  title={Detecting and Reducing the Factual Hallucinations of Large Language Models with Metamorphic Testing},
  author={Wu, Weibin and Cao, Yuhang and Yi, Ning and Ou, Rongyi and Zheng, Zibin},
  journal={Proceedings of the ACM on Software Engineering},
  volume={2},
  number={FSE},
  pages={1432--1453},
  year={2025},
  publisher={ACM New York, NY, USA}
}

@article{schafer2023empirical,
  title={An empirical evaluation of using large language models for automated unit test generation},
  author={Sch{\"a}fer, Max and Nadi, Sarah and Eghbali, Aryaz and Tip, Frank},
  journal={IEEE Transactions on Software Engineering},
  volume={50},
  number={1},
  pages={85--105},
  year={2023},
  publisher={IEEE}
}

@article{jimenez2023swe,
  title={Swe-bench: Can language models resolve real-world github issues?},
  author={Jimenez, Carlos E and Yang, John and Wettig, Alexander and Yao, Shunyu and Pei, Kexin and Press, Ofir and Narasimhan, Karthik},
  journal={arXiv preprint arXiv:2310.06770},
  year={2023}
}

@article{austin2021program,
  title={Program synthesis with large language models},
  author={Austin, Jacob and Odena, Augustus and Nye, Maxwell and Bosma, Maarten and Michalewski, Henryk and Dohan, David and Jiang, Ellen and Cai, Carrie and Terry, Michael and Le, Quoc and others},
  journal={arXiv preprint arXiv:2108.07732},
  year={2021}
}

@article{chen2020metamorphic,
  title={Metamorphic testing: a new approach for generating next test cases},
  author={Chen, Tsong Y and Cheung, Shing C and Yiu, Shiu Ming},
  journal={arXiv preprint arXiv:2002.12543},
  year={2020}
}

@article{ouedraogo2024large,
  title={Large-scale, Independent and Comprehensive study of the power of LLMs for test case generation},
  author={Ou{\'e}draogo, Wendk{\^u}uni C and Kabor{\'e}, Kader and Li, Yinghua and Tian, Haoye and Koyuncu, Anil and Klein, Jacques and Lo, David and Bissyand{\'e}, Tegawend{\'e} F},
  journal={arXiv preprint arXiv:2407.00225},
  year={2024}
}

@inproceedings{reimers-2019-sentence-bert,
    title = "Sentence-BERT: Sentence Embeddings using Siamese BERT-Networks",
    author = "Reimers, Nils and Gurevych, Iryna",
    booktitle = "Proceedings of the 2019 Conference on Empirical Methods in Natural Language Processing",
    month = "11",
    year = "2019",
    publisher = "Association for Computational Linguistics",
    url = "http://arxiv.org/abs/1908.10084",
}

@article{segura2016survey,
  title={A survey on metamorphic testing},
  author={Segura, S. and P{\'e}rez, A. and Cort{\'e}s, A. R. and Polo, M.},
  journal={IEEE Transactions on Software Engineering},
  volume={42},
  number={9},
  pages={805--824},
  year={2016}
}

@inproceedings{jin2024can,
  title={Can chatgpt support developers? an empirical evaluation of large language models for code generation},
  author={Jin, Kailun and Wang, Chung-Yu and Pham, Hung Viet and Hemmati, Hadi},
  booktitle={Proceedings of the 21st International Conference on Mining Software Repositories},
  pages={167--171},
  year={2024}
}

@article{barke2023grounded,
  title={Grounded copilot: How programmers interact with code-generating models},
  author={Barke, Shraddha and James, Michael B and Polikarpova, Nadia},
  journal={Proceedings of the ACM on Programming Languages},
  volume={7},
  number={OOPSLA1},
  pages={85--111},
  year={2023},
  publisher={ACM New York, NY, USA}
}

@article{chen2021evaluating,
  title={Evaluating large language models trained on code},
  author={Chen, Mark and Tworek, Jerry and Jun, Heewoo and Yuan, Qiming and Pinto, Henrique Ponde De Oliveira and Kaplan, Jared and Edwards, Harri and Burda, Yuri and Joseph, Nicholas and Brockman, Greg and others},
  journal={arXiv preprint arXiv:2107.03374},
  year={2021}
}

@inproceedings{chen2024chatunitest,
  title={Chatunitest: A framework for llm-based test generation},
  author={Chen, Yinghao and Hu, Zehao and Zhi, Chen and Han, Junxiao and Deng, Shuiguang and Yin, Jianwei},
  booktitle={Companion Proceedings of the 32nd ACM International Conference on the Foundations of Software Engineering},
  pages={572--576},
  year={2024}
}

@inproceedings{alshahwan2024observation,
  title={Observation-based unit test generation at meta},
  author={Alshahwan, Nadia and Harman, Mark and Marginean, Alexandru and Tal, Rotem and Wang, Eddy},
  booktitle={Companion Proceedings of the 32nd ACM International Conference on the Foundations of Software Engineering},
  pages={173--184},
  year={2024}
}

@inproceedings{xia2023automated,
  title={Automated program repair in the era of large pre-trained language models},
  author={Xia, Chunqiu Steven and Wei, Yuxiang and Zhang, Lingming},
  booktitle={2023 IEEE/ACM 45th International Conference on Software Engineering (ICSE)},
  pages={1482--1494},
  year={2023},
  organization={IEEE}
}

@inproceedings{xu2022systematic,
  title={A systematic evaluation of large language models of code},
  author={Xu, Frank F and Alon, Uri and Neubig, Graham and Hellendoorn, Vincent Josua},
  booktitle={Proceedings of the 6th ACM SIGPLAN international symposium on machine programming},
  pages={1--10},
  year={2022}
}

@article{achiam2023gpt,
  title={Gpt-4 technical report},
  author={Achiam, Josh and Adler, Steven and Agarwal, Sandhini and Ahmad, Lama and Akkaya, Ilge and Aleman, Florencia Leoni and Almeida, Diogo and Altenschmidt, Janko and Altman, Sam and Anadkat, Shyamal and others},
  journal={arXiv preprint arXiv:2303.08774},
  year={2023}
}

@misc{mistral_large,
  title={Mistral Large},
  author={Mistral AI},
  year={2024},
  url={https://mistral.ai/models},
  note={Accessed: 2025-03-08}
}

@article{agarwal2025gpt,
  title={gpt-oss-120b \& gpt-oss-20b model card},
  author={Agarwal, Sandhini and Ahmad, Lama and Ai, Jason and Altman, Sam and Applebaum, Andy and Arbus, Edwin and Arora, Rahul K and Bai, Yu and Baker, Bowen and Bao, Haiming and others},
  journal={arXiv preprint arXiv:2508.10925},
  year={2025}
}

@article{yang2025qwen3,
  title={Qwen3 technical report},
  author={Yang, An and Li, Anfeng and Yang, Baosong and Zhang, Beichen and Hui, Binyuan and Zheng, Bo and Yu, Bowen and Gao, Chang and Huang, Chengen and Lv, Chenxu and others},
  journal={arXiv preprint arXiv:2505.09388},
  year={2025}
}

@article{agentless,
  author    = {Xia, Chunqiu Steven and Deng, Yinlin and Dunn, Soren and Zhang, Lingming},
  title     = {Agentless: Demystifying LLM-based Software Engineering Agents},
  year      = {2024},
  journal   = {arXiv preprint},
}

@article{hoard2025acceptance,
  title={The acceptance, use, and perceptions of metamorphic testing for a sample of open-source software developers},
  author={Hoard, B. R.},
  journal={Cogent Engineering},
  volume={12},
  number={1},
  pages={2522652},
  year={2025},
  publisher={Taylor \& Francis}
}

@inproceedings{liu2012new,
  title={A new method for constructing metamorphic relations},
  author={Liu, Huai and Liu, Xuan and Chen, Tsong Yueh},
  booktitle={2012 12th international conference on quality software},
  pages={59--68},
  year={2012},
  organization={IEEE}
}

@article{li2024metamorphic,
  title={Metamorphic relation generation: State of the art and visions for future research},
  author={Li, Rui and Liu, Huai and Poon, Pak-Lok and Towey, Dave and Sun, Chang-Ai and Zheng, Zheng and Zhou, Zhi Quan and Chen, Tsong Yueh},
  journal={arXiv preprint arXiv:2406.05397},
  year={2024}
}

@inproceedings{xu2024mr,
  title={MR-Adopt: Automatic deduction of input transformation function for metamorphic testing},
  author={Xu, Congying and Chen, Songqiang and Wu, Jiarong and Cheung, Shing-Chi and Terragni, Valerio and Zhu, Hengcheng and Cao, Jialun},
  booktitle={Proceedings of the 39th IEEE/ACM International Conference on Automated Software Engineering},
  pages={557--569},
  year={2024}
}

@article{Chen2024NLPerturbatorST,
  title={NLPerturbator: Studying the Robustness of Code LLMs to Natural Language Variations},
  author={Junkai Chen and Zhenhao Li and Hu Xing and Xia Xin},
  journal={ACM Transactions on Software Engineering and Methodology},
  year={2024},
  url={https://api.semanticscholar.org/CorpusID:270845488}
}

@article{Shen2023InCW,
  title={In ChatGPT We Trust? Measuring and Characterizing the Reliability of ChatGPT},
  author={Xinyue Shen and Zeyuan Chen and Michael Backes and Yang Zhang},
  journal={ArXiv},
  year={2023},
  volume={abs/2304.08979},
  url={https://api.semanticscholar.org/CorpusID:258187122}
}

@article{Liu2024LargeLM,
  title={Large Language Model-Based Agents for Software Engineering: A Survey},
  author={Junwei Liu and Kaixin Wang and Yixuan Chen and Xin Peng and Zhenpeng Chen and Lingming Zhang and Yiling Lou},
  journal={ArXiv},
  year={2024},
  volume={abs/2409.02977},
  url={https://api.semanticscholar.org/CorpusID:272423732}
}

@article{yang2024swe,
  title={Swe-agent: Agent-computer interfaces enable automated software engineering},
  author={Yang, John and Jimenez, Carlos E and Wettig, Alexander and Lieret, Kilian and Yao, Shunyu and Narasimhan, Karthik and Press, Ofir},
  journal={Advances in Neural Information Processing Systems},
  volume={37},
  pages={50528--50652},
  year={2024}
}

@article{wang2024openhands,
  title={Openhands: An open platform for ai software developers as generalist agents},
  author={Wang, Xingyao and Li, Boxuan and Song, Yufan and Xu, Frank F and Tang, Xiangru and Zhuge, Mingchen and Pan, Jiayi and Song, Yueqi and Li, Bowen and Singh, Jaskirat and others},
  journal={arXiv preprint arXiv:2407.16741},
  year={2024}
}

@article{qian2023chatdev,
  title={Chatdev: Communicative agents for software development},
  author={Qian, Chen and Liu, Wei and Liu, Hongzhang and Chen, Nuo and Dang, Yufan and Li, Jiahao and Yang, Cheng and Chen, Weize and Su, Yusheng and Cong, Xin and others},
  journal={arXiv preprint arXiv:2307.07924},
  year={2023}
}

@article{Li2024MetamorphicRG,
  title={Metamorphic Relation Generation: State of the Art and Visions for Future Research},
  author={Rui Li and Huai Liu and Pak-Lok Poon and Dave Towey and Chang-Ai Sun and Zheng Zheng and Zhi Quan Zhou and Tsong Yueh Chen},
  journal={ArXiv},
  year={2024},
  volume={abs/2406.05397},
  url={https://api.semanticscholar.org/CorpusID:270371089}
}

@article{chen2018metamorphic,
  title={Metamorphic testing: A review of challenges and opportunities},
  author={Chen, Tsong Yueh and Kuo, Fei-Ching and Liu, Huai and Poon, Pak-Lok and Towey, Dave and Tse, TH and Zhou, Zhi Quan},
  journal={ACM Computing Surveys (CSUR)},
  volume={51},
  number={1},
  pages={1--27},
  year={2018},
  publisher={ACM New York, NY, USA}
}

@article{Wang2024MeTMaPMT,
  title={MeTMaP: Metamorphic Testing for Detecting False Vector Matching Problems in LLM Augmented Generation},
  author={Guanyu Wang and Yuekang Li and Yi Liu and Gelei Deng and Tianlin Li and Guosheng Xu and Yang Liu and Haoyu Wang and Kailong Wang},
  journal={2024 IEEE/ACM First International Conference on AI Foundation Models and Software Engineering (Forge) Conference Acronym:},
  year={2024},
  pages={12-23},
  url={https://api.semanticscholar.org/CorpusID:267782674}
}

@article{Ayerdi2023GenMorphAG,
  title={GenMorph: Automatically Generating Metamorphic Relations via Genetic Programming},
  author={Jon Ayerdi and Valerio Terragni and Gunel Jahangirova and Aitor Arrieta and Paolo Tonella},
  journal={IEEE Transactions on Software Engineering},
  year={2023},
  volume={50},
  pages={1888-1900},
  url={https://api.semanticscholar.org/CorpusID:266551033}
}

@article{Reddy2025MetamorphicTF,
  title={Metamorphic Testing for Fairness Evaluation in Large Language Models: Identifying Intersectional Bias in LLaMA and GPT},
  author={Harishwar Reddy and Madhusudan Srinivasan and Upulee Kanewala},
  journal={2025 IEEE/ACIS 23rd International Conference on Software Engineering Research, Management and Applications (SERA)},
  year={2025},
  pages={239-246},
  url={https://api.semanticscholar.org/CorpusID:277740988}
}

@inproceedings{Guo2024MORTARMM,
  title={MORTAR: Multi-turn Metamorphic Testing for LLM-based Dialogue Systems},
  author={Guoxiang Guo and Aldeida Aleti and Neelofar Neelofar and Chakkrit Kla Tantithamthavorn and Yuan Yuan Qi and Tsong Yueh Chen},
  year={2024},
  url={https://api.semanticscholar.org/CorpusID:274965380}
}

@inproceedings{Zhang2024CodeAgentEC,
  title={CodeAgent: Enhancing Code Generation with Tool-Integrated Agent Systems for Real-World Repo-level Coding Challenges},
  author={Kechi Zhang and Jia Li and Ge Li and Xianjie Shi and Zhi Jin},
  booktitle={Annual Meeting of the Association for Computational Linguistics},
  year={2024},
  url={https://api.semanticscholar.org/CorpusID:266999556}
}

@article{Wang2023ASO,
  title={A survey on large language model based autonomous agents},
  author={Lei Wang and Chengbang Ma and Xueyang Feng and Zeyu Zhang and Hao-ran Yang and Jingsen Zhang and Zhi-Yang Chen and Jiakai Tang and Xu Chen and Yankai Lin and Wayne Xin Zhao and Zhewei Wei and Ji-rong Wen},
  journal={Frontiers of Computer Science},
  year={2023},
  volume={18},
  url={https://api.semanticscholar.org/CorpusID:261064713}
}

\end{document}